\documentclass[12pt]{article}
\usepackage[pdftex]{color}
\usepackage{cancel} %
\usepackage{amsmath,amsthm,amssymb}
\usepackage{wasysym}
\usepackage{graphicx}
\usepackage{color}%
\usepackage[onehalfspacing]{setspace}
\usepackage{lipsum}
\usepackage[lmargin=2.5cm,rmargin=2.5cm,tmargin=3cm,bmargin=3cm]{geometry}
\usepackage{multirow}
\usepackage{hyperref}
\hypersetup{hidelinks}
\usepackage{cleveref}
\usepackage{accents}
\usepackage[ruled,vlined,linesnumbered]{algorithm2e}
\usepackage[authoryear]{natbib}
\usepackage{lineno}
\usepackage{scalerel}
\usepackage{stackengine}
\usepackage{nomencl}
\renewcommand\nomgroup[1]{%
  \item[\bfseries
  \ifstrequal{#1}{A}{Variables}{%
  \ifstrequal{#1}{B}{Superscripts}{%
  \ifstrequal{#1}{C}{Subscripts}{%
  \ifstrequal{#1}{D}{Abbreviations}{
  \ifstrequal{#1}{E}{Special functions}{
  \ifstrequal{#1}{F}{Operators}{}}}}}}%
]}
\makenomenclature
\stackMath
\definecolor{color}{rgb}{0.8500, 0.3250, 0.0980} 

\newcommand\keywords[1]{\textbf{Keywords}: #1}
\let\oldequation\equation
\let\oldendequation\endequation
\renewenvironment{equation}
{\linenomathNonumbers\oldequation}
{\oldendequation\endlinenomath}

\begin{document}
\begin{center}
{\rm \bf \Large{A stochastic model for elastoplastic contact of rough surfaces incorporating scale-dependent hardness}}

\end{center}

\begin{center}
{\bf Yang Xu}$^{\text{ab}}$\footnote{Corresponding author: yang.xu@hfut.edu.cn}, {\bf Hengxu Song}$^{\text{cd}}$, {\bf Jianqiao Hu}$^{\text{cd}}$\footnote{Corresponding author: jianqiaohu@imech.ac.cn}
\end{center}
\begin{flushleft}
{
$^{\text{a}}$School of Mechanical Engineering, Hefei University of Technology, Hefei, 230009, China \\
$^{\text{b}}$Anhui Province Key Laboratory of Digital Design and Manufacturing, Hefei, 230009, China \\
$^{\text{c}}$LNM, Institute of Mechanics, Chinese Academy of Sciences, Beijing, 100190, China \\
$^{\text{d}}$University of Chinese Academy of Sciences, Beijing, 100049, China
}
\end{flushleft}

\begin{abstract}

The stress concentrations caused by inherent roughness of natural and manufactured surfaces often induce plastic deformation at contact interfaces, a challenge compounded by competing influences of the size effect of plastic deformation and self-affine rough surface topography. To address this, we developed a novel methodology based on stochastic theory using compounded Chapman-Kolmogorov equations, for the first time, to solve elastoplastic contact problems involving scale-dependent hardness. Our approach formulates three integral equations describing the evolution of probability density functions of elastic contact pressure, relative plastic contact area, and relative non-contact area across geometrical scales. We thoroughly investigate the effects of scale-dependent hardness on contact pressure distribution, relative elastic and plastic contact areas, and the area-to-load relationship. By adjusting various mechanical and material properties, our model predicts a smooth transition from linear elasticity to elastic-plastic behavior and finally to full plasticity. A key advancement is the derivation of a new topographic yield parameter incorporating a wider range of material and geometrical properties, aiding identification of contact status. Numerical solutions enable highly precise determination of elastic and plastic limits via curve-fitting, and we also provide a new diagram for rapid identification of contact status. This study pioneers a stochastic process framework for applying the compounded Chapman-Kolmogorov equation to rough surface contact analysis, and the integral equations characterizing how interfacial properties evolve with scale could offer valuable insights for other multidisciplinary fields where multiscale roughness is critical, such as earthquakes, electrical contact, and contact electrification.

\end{abstract}

\keywords{Scale-dependent hardness; Rough surface contact; Elastoplastic contact; Chapman-Kolmogorov equation; Stochastic process}

\nomenclature[A]{$C(q)$}{Power spectral density in terms of wavenumber ($2\pi$ times the reciprocal of the wavelength)}

\nomenclature[A]{$C_0$}{Constant proportionality in the power law of PSD}

\nomenclature[A]{$q_{\text{r}}$}{Roll-off wavenumber}

\nomenclature[A]{$q_{\text{s}}$}{Upper cut-off wavenumber}

\nomenclature[A]{$q_{\text{l}}$}{Lower cut-off wavenumber}

\nomenclature[A]{$H$}{Hurst exponent}

\nomenclature[A]{$\lambda_{\text{s}}$}{Shortest wavelength  of a band-width limited rough surface}

\nomenclature[A]{$\lambda_{\text{l}}$}{Longest wavelength of a band-width limited rough surface}

\nomenclature[A]{$\zeta$}{Magnification $\zeta = \lambda_{\text{l}}/\lambda_{\text{s}} = q_{\text{s}}/q_{\text{l}}$}

\nomenclature[A]{$p(\boldsymbol{x}, \zeta)$}{Contact pressure distribution in the function of in-plane coordinate vector $\boldsymbol{x} = (x, y)$ and $\zeta$}

\nomenclature[A]{$H(\zeta)$}{Scale-dependent hardness}

\nomenclature[A]{$P(p, \zeta)$}{PDF of $p(\boldsymbol{x}, \zeta) = p$}

\nomenclature[A]{$P(p, \zeta|p', \zeta')$}{Transition PDF of $p(\boldsymbol{x}, \zeta) = p$ given $p(\boldsymbol{x}, \zeta') = p'$}

\nomenclature[A]{$H_0$}{Constant hardness or initial hardness at $\zeta = 1$}

\nomenclature[A]{$A_{\text{nc}}^*$}{The ratio of non-contact area to the nominal contact area}

\nomenclature[A]{$A_{\text{el}}^*$}{The ratio of elastic contact area to the nominal contact area}

\nomenclature[A]{$A_{\text{pl}}^*$}{The ratio of plastic contact area to the nominal contact area}

\nomenclature[A]{$A_{\text{pl}}^*\bigg|_{p', \zeta'}$}{Conditional probability of $p(\boldsymbol{x}, \zeta) = H_0$ (constant hardness) or $p(\boldsymbol{x}, \zeta) = H(\zeta)$ (scale-dependent hardness) given $p(\boldsymbol{x}, \zeta') = p'$}

\nomenclature[A]{$A_{\text{nc}}^*\bigg|_{p', \zeta'}$}{Conditional probability of $p(\boldsymbol{x}, \zeta) = 0$ given $p(\boldsymbol{x}, \zeta') = p'$}

\nomenclature[A]{$P_0(p, \zeta)$}{PDF of $p(\boldsymbol{x}, \zeta) = p \in (0, H_0)$ (constant hardness) or $p(\boldsymbol{x}, \zeta) = p \in (0, H(\zeta))$ (scale-dependent hardness)}

\nomenclature[A]{$P_0(p, \zeta|p', \zeta')$}{PDF of $p(\boldsymbol{x}, \zeta) = p \in (0, H_0)$ (constant hardness) or $p(\boldsymbol{x}, \zeta) = p \in (0, H(\zeta))$ (scale-dependent hardness) given $p(\boldsymbol{x}, \zeta') = p'$}

\nomenclature[A]{$V$}{Variance of contact pressure}

\nomenclature[A]{$E^*$}{Plane strain modulus, $E^* = E/(1 - \nu^2)$}

\nomenclature[A]{$C$}{Constant}

\nomenclature[A]{$\Delta V$}{Incremental change of pressure variance when magnification increases from $\zeta'$ to $\zeta$, $\Delta V = V(\zeta) - V(\zeta')$}

\nomenclature[A]{$a', b'$}{Two coefficients in the closed form of $P_0(p, \zeta|p', \zeta')$}

\nomenclature[A]{$H_1$}{Hardness value at magnification $\zeta'$}

\nomenclature[A]{$H_2$}{Hardness value at magnification $\zeta$}

\nomenclature[A]{$\Delta H$}{Incremental change of hardness when magnification increases from $\zeta'$ to $\zeta$, $\Delta H = H(\zeta) - H(\zeta')$}

\nomenclature[A]{$\Delta \zeta$}{Mesh interval of discretized magnification axis using a logarithmic scale.}

\nomenclature[A]{$\Delta p$}{Mesh interval of discretized pressure axis using a linear scale}

\nomenclature[A]{$n$}{Exponent of scale-dependent hardness}

\nomenclature[A]{$\bar{p}$}{Average contact pressure}

\nomenclature[A]{$A^*$}{The ratio of real contact area to the nominal contact area}

\nomenclature[A]{$\kappa$}{Proportionality of load-to-area relation under light load condition}

\nomenclature[A]{$n_{\zeta}$}{Number of discretization points on the magnification axis}

\nomenclature[A]{$n_p$}{Number of pressure-grid points}

\nomenclature[A]{$n_{p0}$}{Reference number of pressure-grid points used to define $\Delta p_0$}

\nomenclature[A]{$E$}{Young's modulus}

\nomenclature[A]{$\nu$}{Poisson's ratio}

\nomenclature[A]{$\sqrt{\langle |\nabla h |^2\rangle}$}{Root mean square gradient}

\nomenclature[A]{$P(h, \zeta)$}{PDF of $h(\boldsymbol{x}, \zeta) = h$}

\nomenclature[A]{$P(h, \zeta|h', \zeta')$}{PDF of $h(\boldsymbol{x}, \zeta) = h$ given $h(\boldsymbol{x}, \zeta') = h'$}

\nomenclature[A]{$\Delta V_h$}{Incremental change of rough surface height variance when magnification increases from $\zeta'$ to $\zeta$, i.e., $\Delta V_h = V_h(\zeta) - V_h(\zeta')$}

\nomenclature[A]{$w$}{Topographic yield parameter}

\nomenclature[A]{$M_{\text{le}}$}{Misfit parameter 1}

\nomenclature[A]{$a, b$}{Two coefficients used in the closed-form solution of $P_0(p, \zeta)$ when hardness is constant}

\nomenclature[A]{$M_{\text{fp}}$}{Misfit parameter 2}

\nomenclature[A]{$V_h$}{Variance of rough surface height}

\nomenclature[D]{PDF}{Probability density function}

\nomenclature[D]{GND}{Geometrically necessary dislocation}

\nomenclature[D]{PSD}{Power spectral density}

\nomenclature[D]{C-K}{Chapman-Kolmogorov}

\nomenclature[E]{$\delta(x)$}{Dirac delta function}

\nomenclature[E]{$\mathcal{H}(x)$}{Heaviside function, where $\mathcal{H}(x > 0) = 1$ and $\mathcal{H}(x < 0) = 0$}

\nomenclature[E]{$\text{erf}(x)$}{Error function}

\nomenclature[E]{$\text{erfc}(x)$}{Complementary error function, $\text{erfc}(x) = 1 - \text{erf}(x)$}

\printnomenclature

\section{Introduction}
Surface roughness reduces the real contact area due to the complex solid-solid interaction between rough surfaces. The stress concentration at the tips of the contact asperities is likely to cause yielding in all metallic materials. Elastoplastic contact between rough surfaces plays an important role in many tribological problems, e.g., electrical contact \citep{jackson2015rough}, metallic seal \citep{fischer2020fluid, perez2016modelling}, and earthquakes \citep{lambert2025competition}. Recent work by \citet{sobarzo2025spontaneous} suggested that contact electrification between identical materials is strongly influenced by plasticity-induced permanent deformation. 

The most popular method for solving elastoplastic rough contact is the multi-asperity contact model \citep{jackson2006statistical} originally developed by \citet{Greenwood66}. Assuming that the load is supported by asperities, the height of which follows a certain distribution, the relationship between the normal load, the real area of contact, and the mean interfacial gap can be formulated by a system of integral equations. An empirical elastoplastic contact model at the asperity level can be curve-fitted from the finite element results \citep{jackson2005finite, ghaednia2017review}. Recently, Wang and his collaborators \citep{liang2022elastic,ding2022incremental, li2023improved} have published a series of works on a novel incremental flat-end punch approach to solving elastoplastic contact problems. The finite element method is the most ready-to-implement numerical approach to solving the elastoplastic contact problem \citep{pei2005finite, wang2018theoretical, an2019deterministic, wang2020effect, zhang2024contact}. In addition to the finite element method, the boundary element method is an alternative approach that enables high resolution of the rough surface topography. Plasticity is introduced in the boundary element method precisely through a complex fundamental solution \citep{jacq2002development,frerot2019fourier} or approximately by truncating the contact pressure to the hardness value, which is commonly approximated as three times the yield strength \citep{almqvist2007dry,sahlin2010mixed}. Both finite element and boundary element methods suffer serious drawbacks, as the computational time significantly increases with the increased resolution of point clouds. 

The aforementioned methods are known as the single-scale model due to the fact that the input point cloud data or the height distribution function are obtained by sampling the rough surface topography with a fixed sampling resolution. \citet{majumdar1990} proposed a multiscale contact model in which the surface is characterized using the Weierstrass-Mandelbrot function, and the plasticity is approximated using a constant hardness model. \citet{Persson01b} developed a diffusion equation to characterize the evolution of the probability density function (PDF) of contact pressure with scale. The upper limit of contact pressure is constrained by a constant hardness. The closed form of the pressure PDF and the real area of contact have been obtained as a Fourier series. \citet{Xu22} further simplified the Fourier series solution to the superposition of three Gaussian distributions. In the same year, two independent multiscale elastoplastic contact models \citep{jackson2006multi,gao2006elastic} were developed based on the concept of ``protuberance-on-protuberance" originally proposed by \citep{archard1957elastic}. These two models are discussed in depth later in Section \ref{subsec:model_discussion}.  

Hardness, defined as the mean pressure on the plastically deformed area, is not an intrinsic property of the elastoplastic material. \citet{jackson2005finite} pointed out that the hardness of an elastoplastic sphere decreases with increasing indentation depth. \citet{jackson2015solution} derived a closed-form solution for the hardness-to-yield strength ratio for plane contact using slip line theory, which is in good agreement with the finite element results. The results of the nano-indentation test \citep{swadener2002correlation} indicate that hardness varies with the size of the indenter. This size effect is due to the fact that Geometrically Necessary Dislocations (GNDs) accumulate at the valley of the indentation zone to accommodate the high strain gradient. The increase in the density of GNDs leads to a strengthening effect on the yield strength \citep{nix1998indentation}, which is known as strain gradient plasticity. 

The strain gradient plasticity was initially implemented in the multi-asperity contact model to account for the scale-dependent hardness \citep{bhushan2003scale, jackson2006effect,song2017statistical}, where the hardness at the single asperity level is determined either analytically by the Nix-Gao model \citep{jackson2006effect} or numerically using the finite element method \citep{song2017statistical}. \citet{song2016strain} developed a finite element rough surface contact model in which conventional mechanism-based strain gradient plasticity was adopted to account for scale-dependent hardness. The yield strength, as well as the hardness, can be increased by manipulating the empirical parameter (intrinsic length). The PDF of the contact pressure is broadened over a larger pressure range as the hardness increases. \citet{venugopalan2019plastic} developed a discrete dislocation dynamics model to simulate an elastoplastic half-plane in purely normal contact with a rigid flat. The PDFs of the contact pressure predicted by the dislocation dynamics model and Persson's theory \citep{Persson01b} agree well when the normal load and the root mean square slope are relatively low. \citet{venugopalan2017green} found that the accuracy of Persson's theory can be improved if the upper pressure limit obtained from the numerical model is used. However, the deviation is still too obvious to ignore. \citet{persson2006contact} extended his elastoplastic contact theory to include the scale-dependent hardness, and \citet{lambert2025competition} developed the numerical model for solving it until recently. The evolution of the PDF of the elastic contact pressure with respect to scale is characterized by a diffusion equation with a moving boundary condition governed by scale-dependent hardness. A closed-form topographic yield parameter has been found to characterize the competition between the broadening of the PDF of the elastic contact pressure and the increase of hardness with scale. Similar attempts have been made by \citet{ciavarella2024new} when the contact is in the fully plastic limit. \citet{jackson2023asperity} developed two multiscale elastoplastic models with a scale-dependent yield strength. One is based on the multi-asperity contact model, where moments of rough surface height are scale-dependent, and the yield strength used in the elastoplastic asperity contact model is estimated by the Nix and Gao theory. The other is an extension of the Jackson-Streator model \citep{jackson2006multi} in which the scale-dependent yield strength is implemented in the sinusoidal waviness contact model. 

Inspired by the work of \citet{lambert2025competition, gao2006elastic, jackson2006multi}, we extend Persson's theory of contact to simulate elastoplastic rough surface contact with scale-dependent hardness by assuming that the variation of the random contact pressure with respect to scale is a Markov process. The elastoplastic contact problem is introduced in Section 2. In Section 3, the Chapman-Kolmogorov equation for elastoplastic contact with constant hardness is derived. In Section 4, this formulation is further extended to incorporate scale-dependent hardness. The numerical implementation of the present model is explained in Section 5. Some representative results are provided in Section 6. In Section 7, similar multiscale models and the governing equations from the work of \citep{lambert2025competition} are discussed in detail, followed by a new topographic yield parameter and the elastoplastic contact status diagram.

\section{Problem statement}
\begin{figure}[h!]
  \centering
  \includegraphics[width=10cm]{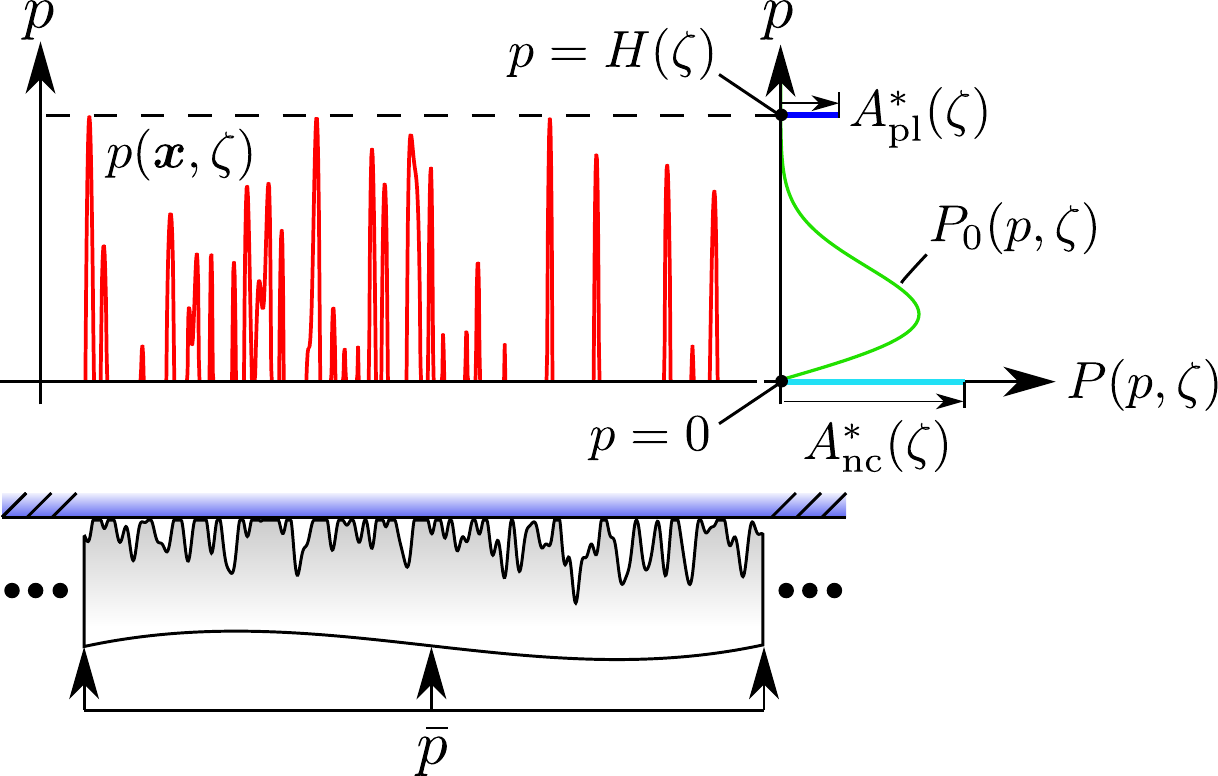}
  \caption{A cross-sectional view of the purely normal contact between a rigid flat and a rough surface and the schematic of the probability density function of the contact pressure.}\label{fig:Fig_1}
\end{figure}

In the present study, we examine the purely normal elastoplastic contact between a rigid flat and a half-space, see Fig. \ref{fig:Fig_1}. The boundary of the half-space is a bandwidth-limited, nominally flat, isotropic, and self-affine rough surface. The power spectral density (PSD) of the rough surface is characterized by the following power law \citep{Jacobs17, pradhan2025surface}: 
\begin{equation}\label{eq:PSD_piecewise}
C(q) = 
\begin{cases}
C_0 q_{\text{r}}^{-2(1 + H)} ~~~ q \in [q_{\text{l}}, \min(q_{\text{r}}, q_{\text{s}})), \\
C_0 q^{-2(1 + H)} ~~~ q \in [\min(q_{\text{r}}, q_{\text{s}}), q_{\text{s}}], \\
0 ~~~ \text{elsewhere},
\end{cases}
\end{equation}
where $q_{\text{l}} = 2 \pi/\lambda_{\text{l}}$, $q_{\text{r}}$, and $q_{\text{s}} = 2 \pi /\lambda_{\text{s}}$ are the lower cut-off, roll-off, and upper cut-off wavenumbers, respectively; $\lambda_{\text{l}}$ and $\lambda_{\text{s}}$ are the longest and shortest wavelengths, respectively. The Hurst exponent $H \in [0, 1]$; $C_0$ is a constant. Let us define the magnification as $\zeta = q_{\text{s}}/q_{\text{l}}$, which describes the scale at which the rough surface topography is observed. For a given $\zeta$, the contact pressure distribution is denoted by $p(\boldsymbol{x}, \zeta)$. Plasticity is introduced by truncating the contact pressure at the upper limit represented by the scale-dependent hardness $H(\zeta)$, where strain hardening is excluded (Fig. \ref{fig:Fig_1}). Therefore, within the elastic and plastic contact areas, contact pressure satisfies $p(\boldsymbol{x}, \zeta) < H(\zeta)$ and $p(\boldsymbol{x}, \zeta) = H(\zeta)$, respectively. 

\section{Chapman-Kolmogorov equation: constant hardness}\label{sec:CK_constant}
Let us revisit the elastoplastic contact theory with constant hardness $H_0$ initially proposed by \citet{Persson01b}. \citet{xu2024persson} show that the variation of the random elastic contact pressure with the scale follows a Markov process. Assuming that this Markovian nature still holds under elastoplastic contact, the variation of the PDF of $p$ between two neighboring magnifications ($P(p, \zeta)$ and $P(p', \zeta' < \zeta)$) is characterized by the Chapman-Kolmogorov equation:
\begin{equation}\label{eq:CK_H0}
P(p, \zeta) = \int_0^{\infty} P(p, \zeta|p', \zeta') P(p', \zeta') dp',
\end{equation}
where $P(p, \zeta|p', \zeta')$  is the transition probability that represents the conditional probability of $p(\boldsymbol{x}, \zeta) = p$ given $p(\boldsymbol{x}, \zeta') = p'$. Since $P(p, \zeta)$ possesses distinct characteristics in non-contact ($p = 0$), elastic contact ($p \in (0, H_0)$), and plastic contact ($p = H_0$) regions (see Fig. \ref{fig:Fig_1} where $H(\zeta) = H_0$), it can be further expanded in composite form as follows: 
\begin{equation}\label{eq:P_composite}
P(p, \zeta) = A_{\text{nc}}^*(\zeta) \delta(p) + P_0(p, \zeta) \left[\mathcal{H}(p) - \mathcal{H}(p - H_0) \right] + A_{\text{pl}}^*(\zeta) \delta(p - H_0),
\end{equation}
where $A_{\text{nc}}^*$ and $A_{\text{pl}}^*$ represent the fractions of the nominal contact area that are out of contact and in plastic contact, respectively. $P_0(p \in (0, H(\zeta)), \zeta)$ denotes the PDF of elastic contact. $\mathcal{H}(x)$ is the Heaviside function, where $\mathcal{H}(x > 0) = 1$ and $\mathcal{H}(x < 0) = 0$. Similarly, the transition probability can be decomposed in the same manner:
\begin{equation}\label{eq:P_transition}
P(p, \zeta|p', \zeta') = A_{\text{nc}}^*(\zeta)\bigg|_{p', \zeta'} \delta(p) + P_0(p, \zeta|p', \zeta') \left[ \mathcal{H}(p) - \mathcal{H}(p - H_0) \right] + A_{\text{pl}}^*(\zeta)\bigg|_{p', \zeta'} \delta(p - H_0),
\end{equation}
where $A_{\text{nc}}(\zeta)\bigg|_{p', \zeta'}$, $A_{\text{pl}}(\zeta)\bigg|_{p', \zeta'}$, and $P_0(p, \zeta|p', \zeta')$ represent the conditional probabilities of $p(\boldsymbol{x}, \zeta) = 0$, $p(\boldsymbol{x}, \zeta) = H_0$, and $p(\boldsymbol{x}, \zeta) = p \in (0, H_0)$ given $p(\boldsymbol{x}, \zeta') = p' \in [0, H_0]$, respectively. The transition PDF serves as a bridge connecting two pressure PDFs at two different magnifications. 

Since $p \geq 0$, the conservation of probability requires the following identity:
\begin{equation}\label{eq:probability_conservation}
\int_{0}^{H_0} P_0(p, \zeta|p', \zeta') dp +  A_{\text{pl}}^*(\zeta) \bigg|_{p', \zeta'} +  A_{\text{nc}}^*(\zeta) \bigg|_{p', \zeta'} = 1.
\end{equation}
Expanding all PDFs in Eq. \eqref{eq:CK_H0} using their composite forms (Eqs. \eqref{eq:P_composite} and \eqref{eq:P_transition}), applying $\int_0^{\infty} p \cdots dp$ to both sides of Eq. \eqref{eq:CK_H0}, and using the load equilibrium $\int_0^{\infty} p P(p, \zeta) dp = \int_0^{\infty} p' P(p', \zeta') dp'$, the following identity is derived:
\begin{equation}\label{eq:load_equilibrium1}
\int_{0}^{H_0} p P_0(p, \zeta|p', \zeta') \text{d}p + H_0 \cdot A_{\text{pl}}^*(\zeta) \bigg|_{p', \zeta'} = p',
\end{equation}
where $p' \in [0, H_0]$. Considering the limiting case where $p' = 0$ and taking into account that $P_0(p, \zeta|p', \zeta')$ and $A_{\text{pl}}^*(\zeta)$ are non-negative, we can easily deduce that
\begin{equation}\label{eq:Asymptotic_0}
P_0(p, \zeta|0, \zeta') = 0, ~~~ A_{\text{pl}}^*(\zeta) \bigg|_{0, \zeta'} = 0, ~~~
A_{\text{nc}}^*(\zeta) \bigg|_{0, \zeta'} = 1.
\end{equation}
Eq. \eqref{eq:Asymptotic_0} is the no re-entry assumption adopted in Persson's theory \citep{prodanov2014contact, xu2024persson}, which states that the surface points with $p(\boldsymbol{x}, \zeta') = 0$ can only remain out of contact at any higher magnification, i.e., $p(\boldsymbol{x}, \zeta \geq \zeta') = 0$. Similarly, when $p' = H_0$, the non-negative $P_0(p, \zeta|p', \zeta')$ and $A_{\text{pl}}^*(\zeta)$ lead to the following identity: 
\begin{equation}\label{eq:Asymptotic_H0}
P_0(p, \zeta|H_0, \zeta') = 0, ~~~ 
A_{\text{pl}}^*(\zeta) \bigg|_{H_0, \zeta'} = 1, ~~~
A_{\text{nc}}^*(\zeta) \bigg|_{H_0, \zeta'} = 0.
\end{equation}
Eq. \eqref{eq:Asymptotic_H0} represents an alternative no re-entry assumption that is implicitly assumed in Persson's elastoplastic contact theory \citep{Persson01b, Xu22}. This theory states that the surface points with $p(\boldsymbol{x}, \zeta') = H_0$ remain in plastic contact at an arbitrarily higher magnification, i.e., $p(\boldsymbol{x}, \zeta \geq \zeta') = H_0$. 

Three integral equations are derived if all pressure PDFs are expanded in Eq. \eqref{eq:CK_H0} using their composite forms and no re-entry assumptions (Eqs. \eqref{eq:Asymptotic_0} and \eqref{eq:Asymptotic_H0}): 
\begin{align}
P_0(p, \zeta) = &\int_0^{H_0} P_0(p, \zeta|p', \zeta') P_0(p', \zeta') dp', \label{eq:CK_equ1} \\
A_{\text{pl}}^*(\zeta) =& \int_0^{H_0} A_{\text{pl}}^*(\zeta) \bigg|_{p', \zeta'} P_0(p', \zeta') dp' + A_{\text{pl}}^*(\zeta'), \label{eq:CK_equ2}  \\
A_{\text{nc}}^*(\zeta) =& \int_0^{H_0} A_{\text{nc}}^*(\zeta) \bigg|_{p', \zeta'} P_0(p', \zeta') dp' + A_{\text{nc}}^*(\zeta'). \label{eq:CK_equ3} 
\end{align}
Unlike the diffusion equation in Persson's elastoplastic contact theory \citep{Persson01b}, this is the first time that the elastoplastic contact problem has been formulated using integral equations. Because of $0 \leq \int_0^{H_0} P_0(p, \zeta > \zeta'|p', \zeta') dp < 1$, Eq. \eqref{eq:CK_equ1} implies that $\int_0^{H_0} P_0(p, \zeta) dp$ decreases monotonically with the increase of $\zeta$. The ``leakage" of $P_0(p, \zeta)$ ``flows" to the non-contact and plastic contact regions and results in a monotonically increasing $A_{\text{pl}}^*(\zeta')$ and $A_{\text{nc}}^*(\zeta')$, governed by Eqs. \eqref{eq:CK_equ2} and \eqref{eq:CK_equ3}, respectively.

Now, we propose an approximate approach for finding the closed-form solution of $P_0(p, \zeta|p', \zeta')$. We first adopt the assumption made by \citet{Persson01b} that $P_0(p, \zeta)$ satisfies the following diffusion equation: 
\begin{equation}\label{eq:Diffusion_equ_P0}
\frac{\partial}{\partial \zeta} P_0(p, \zeta) = \frac{1}{2} \frac{\partial V}{\partial \zeta} \frac{\partial^2}{\partial p^2} P_0(p, \zeta), 
\end{equation}
where $\partial V(\zeta)/\partial \zeta$ denotes the rate of change of the variance of the elastic contact pressure with respect to $\zeta$, which is approximated in the present work by a closed-form solution when the rough surface is completely flattened elastically:
\begin{equation}
V(\zeta)= \frac{\pi}{2} (E^*)^2 \int_{q_l}^{\zeta q_l} q^3 C(q) dq, 
\end{equation}
where $E^* = E/(1 - \nu^2)$ with $E$ and $\nu$ being Young's modulus and Poisson's ratio, respectively. Referring to Eq. \eqref{eq:CK_equ1}, we can deduce that the transition probability $P_0(p, \zeta|p', \zeta')$ satisfies the same diffusion equation
\begin{equation}\label{eq:Diffusion_equ_transition}
\frac{\partial}{\partial \zeta} P_0(p, \zeta|p', \zeta') = \frac{1}{2} \frac{\partial V}{\partial \zeta} \frac{\partial^2}{\partial p^2} P_0(p, \zeta|p', \zeta'). 
\end{equation}
The initial condition of Eq. \eqref{eq:Diffusion_equ_transition} is
\begin{equation}\label{eq:ini_cond_transition}
P_0(p, \zeta = \zeta'|p', \zeta') = \delta(p - p'). 
\end{equation}
The boundary conditions at $p = 0, H_0$ can be derived from the zero and first order moments of $P_0(p, \zeta|p', \zeta')$. Applying $\int_0^{H_0} \cdots dp$ to Eq. \eqref{eq:Diffusion_equ_transition} and using the identity Eq. \eqref{eq:probability_conservation}, 
\begin{equation}\label{eq:transition_PDF_conserve_1}
\frac{\partial}{\partial \zeta} A_{\text{pl}}^*(\zeta) \bigg|_{p', \zeta'} = -\frac{1}{2} \frac{\partial V}{\partial \zeta} \frac{\partial}{ \partial p} P_0(p, \zeta|p', \zeta') \bigg|_{H_0}. 
\end{equation}
Applying $\int_0^{H_0} p \cdots dp$ to Eq. \eqref{eq:Diffusion_equ_transition} and using identities (Eqs. \eqref{eq:load_equilibrium1} and \eqref{eq:transition_PDF_conserve_1}) requires the following boundary condition to be satisfied:
\begin{equation}
P_0(0, \zeta|p', \zeta') = P_0(H_0, \zeta|p', \zeta') = C \geq 0,
\end{equation}
which satisfies the diffusion equation and the initial condition. $C$ is a constant. The closed-form solution can be expressed in a triple Gaussian distribution form
\begin{equation}\label{eq:P0p_form_transit}
P_0(p, \zeta |p', \zeta') = \frac{1}{\sqrt{2 \pi \Delta V}} \left\{ \exp \left[ -\frac{(p - p')^2}{2 \Delta V} \right] - a' \exp \left[ -\frac{(p + p')^2}{2 \Delta V} \right] - b' \exp \left[ -\frac{(p - 2 H_0 + p')^2}{2 \Delta V} \right] \right\}, 
\end{equation}
where $\Delta V = V(\zeta) - V(\zeta')$. Following the proof given by \citet{Xu22}, we can rigorously prove that $C = 0$, i.e., $P_0(p, \zeta|p', \zeta')$ has absorbing boundary conditions at $p = 0$ and $p = H_0$. To enforce the absorbing boundary conditions, two unknowns ($a'$ and $b'$) are derived:
\begin{align}
a' &= \left\{ 1 - \exp[-2(H_0 - p') H_0/\Delta V ]\right\}/\left[ 1 - \exp(-2H_0^2/\Delta V)\right], \\
b' &= 1 - a' \cdot \text{exp}\left(-2 p' H_0/\Delta V \right).
\end{align}
It is tedious to solve $A_{\text{pl}}^*(\zeta)$ and $A_{\text{nc}}^*(\zeta)$ from Eqs. \eqref{eq:CK_equ2} and \eqref{eq:CK_equ3}, as it requires finding the closed forms of $A_{\text{nc}}^*(\zeta) \bigg|_{p', \zeta'}$ and $A_{\text{pl}}^*(\zeta) \bigg|_{p', \zeta'}$ first. A convenient approach is to solve it directly using load equilibrium and probability conservation:
\begin{align}
A_{\text{pl}}^*(\zeta) &= \left[ \bar{p} - \int_0^{H_0} p P_0(p, \zeta) dp \right]/H_0, \label{eq:Apl_form} \\
A_{\text{nc}}^*(\zeta) &= 1 - \int_0^{H_0} P_0(p, \zeta) dp - A_{\text{pl}}^*(\zeta). \label{eq:Anc_form}
\end{align}

Considering a monotonically increasing sequence of magnifications $[\zeta_0, \zeta_1, \cdots, \zeta_i, \cdots, \zeta_n, \zeta]$, where $\zeta_0 = 1$, $P(p, \zeta)$ at an arbitrary magnification can be formulated using the following compounded Chapman-Kolmogorov equation:
\begin{equation}
P(p, \zeta) = \int_0^{H_0} P(p, \zeta|p_n, \zeta_n) \int_0^{H_0} P(p_n, \zeta_n|p_{n-1}, \zeta_{n-1}) \cdots  \int_0^{H_0} P(p_1, \zeta_1|p_0, \zeta_0) P(p_0, \zeta_0) dp_0 \cdots dp_{n-1} dp_n. 
\end{equation}
The compounded Chapman-Kolmogorov equation is a concise form that is equivalent to solving Eqs. \eqref{eq:CK_equ1}, \eqref{eq:Apl_form}, and \eqref{eq:Anc_form} sequentially. 

\begin{figure}[h!]
  \centering
  \includegraphics[width=16cm]{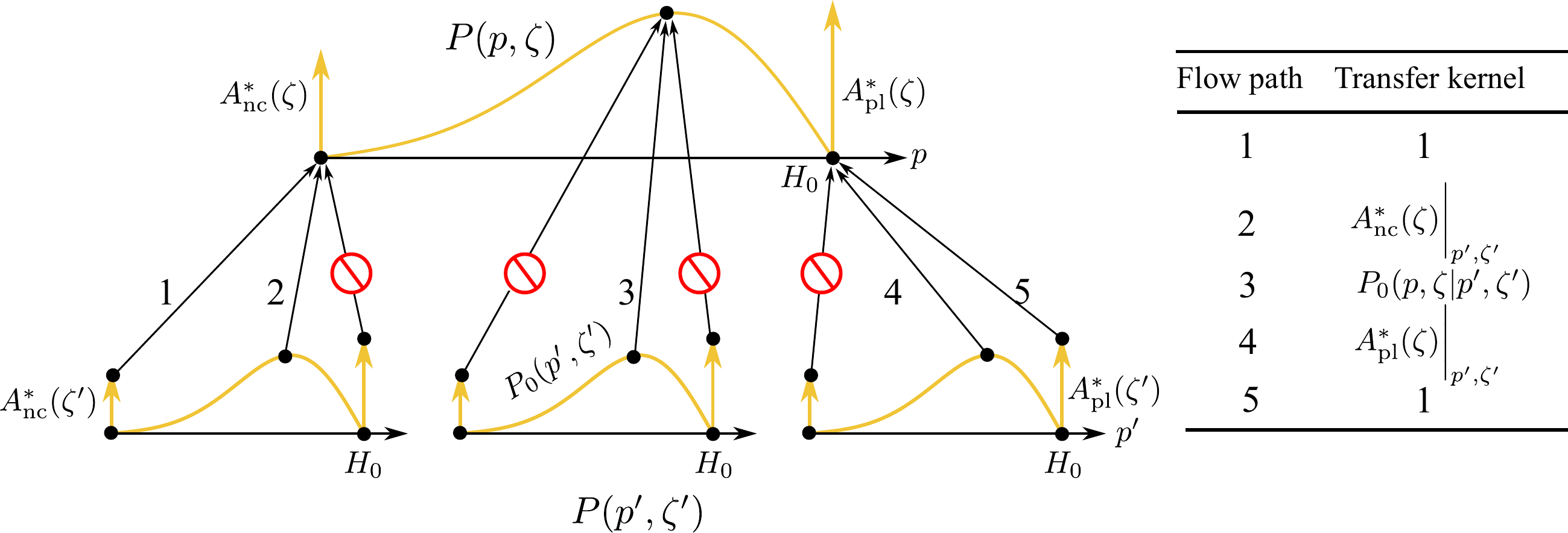}
  \caption{A graphical illustration of probability flow from $P(p', \zeta')$ to $P(p, \zeta)$ where hardness is a constant. The three PDFs at the bottom are all replicas of $P(p', \zeta').$}\label{fig:Fig_2}
\end{figure}

{\bf Remarks} Equations \eqref{eq:CK_equ1}, \eqref{eq:CK_equ2}, and \eqref{eq:CK_equ3} implicitly express the probability flow from $P(p, \zeta)$ to $P(p, \zeta')$, which is graphically illustrated in Fig. \ref{fig:Fig_2}. Theoretically, there are nine paths that facilitate the flow of contact pressure probability from three contact regions: the non-contact area, the elastic contact area, and the plastic contact area, at magnification $\zeta'$ to the same three contact regions at magnification $\zeta$. According to Eqs. \eqref{eq:Asymptotic_0} and \eqref{eq:Asymptotic_H0}, four flow paths are blocked in the present model, as well as Persson's theory of contact, due to the no re-entry assumptions at the non-contact and plastic contact areas:
\begin{itemize}
\item 
$p(\boldsymbol{x}, \zeta') = 0 \not\rightarrow p(\boldsymbol{x}, \zeta) \in H_0$;
\item 
$p(\boldsymbol{x}, \zeta') = 0 \not\rightarrow p(\boldsymbol{x}, \zeta) \in (0, H_0)$;
\item 
$p(\boldsymbol{x}, \zeta') = H_0 \not\rightarrow p(\boldsymbol{x}, \zeta) = 0$;
\item 
$p(\boldsymbol{x}, \zeta') = H_0 \not\rightarrow p(\boldsymbol{x}, \zeta) \in (0, H_0)$.
\end{itemize}
The no re-entry assumption in elastoplastic contact can be articulated as follows: \emph{Given an arbitrary surface point in either the non-contact or plastic contact area, it cannot re-enter other types of areas at any larger magnifications}. This assumption is not accurate in practice. If points within the elastic contact area are allowed to transition to other types of contact areas as $\zeta$ increases, then points within the plastic contact area should similarly be permitted to ``escape". The no re-entry assumption significantly ``drains" the probability of elastic contact pressure, leading to an overestimation of $A_{\text{nc}}^*$ and $A_{\text{pl}}^*$. The reason for this overestimation is a lack of mechanisms escaping the plastic contact area, e.g., strain-hardening law and strain gradient plasticity. 

\section{Chapman-Kolmogorov equation: scale-dependent hardness}\label{sec:CK_mag_hardness}

When hardness becomes scale-dependent, hardness values at two magnifications $\zeta'$ and $\zeta$ are distinguished by $H_1 = H(\zeta')$ and $H_2 = H(\zeta)$, respectively. The composite forms of  $P(p, \zeta)$, $P(p', \zeta')$, and $P(p, \zeta|p', \zeta')$ are slightly different from those presented in Section \ref{sec:CK_constant}: 
\begin{align}
P(p, \zeta) &= A_{\text{nc}}^*(\zeta) \delta(p) + P_0(p, \zeta) \left[\mathcal{H}(p) - \mathcal{H}(p - H_2) \right] + A_{\text{pl}}^*(\zeta) \delta(p - H_2), \label{eq:P0_past_varying}\\
P(p', \zeta') &= A_{\text{nc}}^*(\zeta') \delta(p') + P_0(p', \zeta') \left[\mathcal{H}(p') - \mathcal{H}(p' - H_1) \right] + A_{\text{pl}}^*(\zeta') \delta(p'- H_1), \label{eq:P0_future_varying}\\
P(p, \zeta|p', \zeta') &= A_{\text{nc}}^*(\zeta)\bigg|_{p', \zeta'} \delta(p) + P_0(p, \zeta|p', \zeta') \left[\mathcal{H}(p) - \mathcal{H}(p - H_2) \right] +  A_{\text{pl}}^*(\zeta)\bigg|_{p', \zeta'} \delta(p - H_2). \label{eq:P0_transit_varying}
\end{align}
Substituting Eq. \eqref{eq:P0_transit_varying} into Eq. \eqref{eq:CK_H0}, applying $\int_0^{\infty} p \cdots dp$ to both sides of Eq. \eqref{eq:CK_H0}, and using the load equilibrium $\int_0^{\infty} p P(p, \zeta) dp = \int_0^{\infty} p' P(p', \zeta') dp'$, an identity similar to Eq. \eqref{eq:load_equilibrium1} is established as follows:
\begin{equation}\label{eq:load_equilibrium2}
\int_{0}^{H_2} p P_0(p, \zeta|p', \zeta') \text{d}p + H_2 \cdot A_{\text{pl}}^*(\zeta) \bigg|_{p', \zeta'} = p',
\end{equation}
where $p' \in [0, H_1]$. Equation \eqref{eq:load_equilibrium2} implies that the no re-entry assumption, given in Eq. \eqref{eq:Asymptotic_0}, remains valid. Let $p' = H_1$ in Eq. \eqref{eq:load_equilibrium2}, 
\begin{equation}\label{eq:load_equilibrium3}
\int_{0}^{H_2} p P_0(p, \zeta|H_1, \zeta') \text{d}p + H_2 \cdot A_{\text{pl}}^*(\zeta) \bigg|_{H_1, \zeta'} = H_1.
\end{equation}
The upper limit of the left side of Eq. \eqref{eq:load_equilibrium3} is $H_2$ when $A_{\text{pl}}^*(\zeta) \bigg|_{H_1, \zeta'} = 1$ and $P_0(p, \zeta|H_1, \zeta') = 0$. If $H_1 > H_2$, i.e., $H(\zeta)$ exhibits a decreasing trend against the magnification, Eq. \eqref{eq:load_equilibrium3} cannot be strictly satisfied. Consequently, the present model allows only for an increasing or constant trend of $H(\zeta)$. This requirement is consistent with the size effect in plastic deformation and has been previously established in Persson's theory of contact by \citet{lambert2025competition}, based on the diffusion equation of $P_0(p, \zeta)$. If $H_1 < H_2$, the no re-entry assumption expressed in Eq. \eqref{eq:Asymptotic_H0} is no longer valid. This suggests that surface points with $p(\boldsymbol{x}, \zeta') = H_1$ may enter the elastic contact area or even the non-contact area at a higher magnification $\zeta > \zeta'$. This backflow feature slows down the decreasing rate of $P_0(p, \zeta)$ with respect to $\zeta$. As a matter of fact, the elastoplastic contact with scale-dependent hardness results in a more elastic contact than its counterpart with constant hardness. 

\begin{figure}[h!]
  \centering
  \includegraphics[width=16cm]{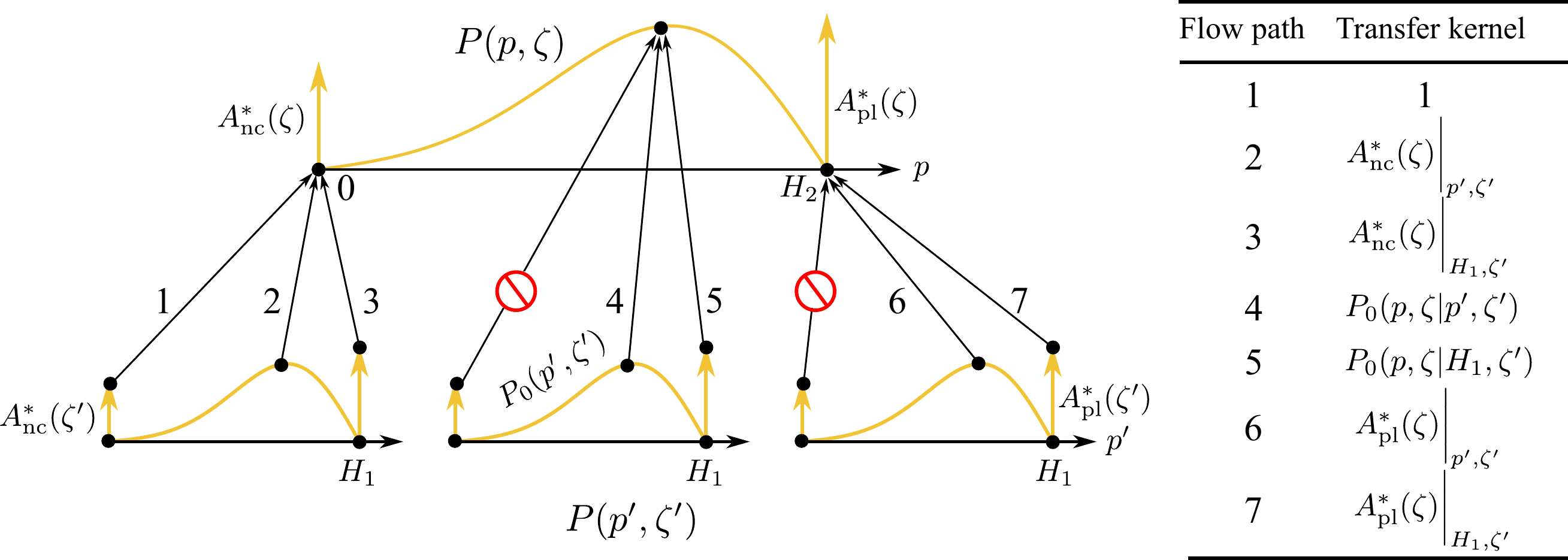}
  \caption{A graphical illustration of probability flow from $P(p', \zeta')$ to $P(p, \zeta)$ where hardness is scale-dependent. The three PDFs at bottom are all replicas of $P(p', \zeta')$}\label{fig:Fig_3}
\end{figure}

Substituting Eqs. \eqref{eq:P0_past_varying}, \eqref{eq:P0_future_varying}, \eqref{eq:P0_transit_varying}, \eqref{eq:Asymptotic_0} into Eq. \eqref{eq:CK_H0}, 
\begin{align}
P_0(p, \zeta) = &\int_0^{H_1} P_0(p, \zeta|p', \zeta') P_0(p', \zeta') dp' + P_0(p, \zeta|H_1, \zeta') A_{\text{pl}}^*(\zeta'), \label{eq:CK_vary_H_P0} \\
A_{\text{pl}}^*(\zeta) =& \int_0^{H_1} A_{\text{pl}}^*(\zeta) \bigg|_{p', \zeta'} P_0(p', \zeta') dp' + A_{\text{pl}}^*(\zeta)\bigg|_{H_1, \zeta'} A_{\text{pl}}^*(\zeta'), \label{eq:CK_vary_H_Apl} \\
A_{\text{nc}}^*(\zeta) =& \int_0^{H_1} A_{\text{nc}}^*(\zeta) \bigg|_{p', \zeta'} P_0(p', \zeta') dp' + A_{\text{nc}}^*(\zeta)\bigg|_{H_1, \zeta'} A_{\text{pl}}^*(\zeta') + A_{\text{nc}}^*(\zeta'). \label{eq:CK_vary_H_Anc} 
\end{align}
Comparing with Eq. \eqref{eq:CK_equ1}, Eq. \eqref{eq:CK_vary_H_P0} includes an additional term on the right side that represents the backflow of pressure probability from the plastic contact area to the elastic contact area as magnification increases from $\zeta'$ to $\zeta$. Similarly, the second term on the right side of Eq. \eqref{eq:CK_vary_H_Anc} indicates that the backflow of pressure probability from the plastic contact area to the non-contact area is also allowed. Therefore, two out of the four blocked paths in elastoplastic contact with constant hardness are now reopened in the present model (see Fig. \ref{fig:Fig_3} for a graphical illustration). Because of $A_{\text{pl}}^*(\zeta) \bigg|_{H_1, \zeta'} < 1$ and $A_{\text{nc}}^*(\zeta) \bigg|_{H_1, \zeta'} < 1$, $A_{\text{pl}}^*(\zeta)$ and $A_{\text{nc}}^*(\zeta)$ may drop with the increasing $\zeta$. This decreasing trend cannot occur when hardness is constant. \citet{lambert2025competition} have confirmed this decreasing trend using the numerical results of Persson's theory of contact, while Eqs. \eqref{eq:CK_vary_H_Apl} and \eqref{eq:CK_vary_H_Anc} provide a mathematical explanation. 

We adopt the assumption made by \citet{lambert2025competition} that $P_0(p, \zeta)$ satisfies the diffusion equation (Eq. \eqref{eq:Diffusion_equ_P0}). Thus, Eq. \eqref{eq:CK_vary_H_P0} implies that the transition probability $P_0(p, \zeta|p' \in (0, H_1], \zeta')$ satisfies the same diffusion equation (Eq. \eqref{eq:Diffusion_equ_transition})\footnote{Since the upper limit $H_1 = H(\zeta')$ is independent of $\zeta$ and $p$, it is treated as a constant when differentiating Eq. \eqref{eq:CK_vary_H_P0} with respect to $\zeta$ and $p$}. Let $\zeta - \zeta'$ be sufficiently small, we can approximate $\partial V/\partial \zeta$ with $\Delta V/\Delta \zeta$, where $\Delta V = V(\zeta) - V(\zeta')$ and Eq. \eqref{eq:Diffusion_equ_transition} can be rewritten as 
\begin{equation}\label{eq:Diffusion_equ_transition_1}
\frac{\partial}{\partial \zeta} P_0(p, \zeta|p', \zeta') = \frac{1}{2} \frac{\Delta V}{\Delta \zeta} \frac{\partial^2}{\partial p^2} P_0(p, \zeta|p', \zeta'). 
\end{equation}

The initial condition remains unchanged from that in Eq. \eqref{eq:ini_cond_transition}. If we assume that the boundary condition at $p = 0$ is an absorbing boundary condition $P_0(p=0, \zeta|p', \zeta') = 0$, the load equilibrium and probability conservation lead to the following non-absorbing boundary condition at $p = H_2$:
\begin{equation} \label{eq:transition_nonabsorbing_BC}
P_0(p=H_2, \zeta|p', \zeta') = \frac{2 \Delta H}{\Delta V} A_{\text{pl}}^*(\zeta)\bigg|_{p', \zeta'},
\end{equation}
where $\Delta H = H_2 - H_1$. A complete derivation of Eq. \eqref{eq:transition_nonabsorbing_BC} is provided in Appendix A. Let $p = H_2$ in Eq. \eqref{eq:CK_vary_H_P0}, the non-absorbing boundary condition of $P_0(p, \zeta)$, originally derived by \citet{lambert2025competition}, is recovered: 
\begin{equation}
P_0(H_2, \zeta) = \frac{2 \Delta H}{\Delta V} A_{\text{pl}}^*(\zeta). 
\end{equation}
The solution of $P_0(p, \zeta|p', \zeta')$ is nearly identical to that in Eq. \eqref{eq:P0p_form_transit}:
\begin{equation}\label{eq:P0p_transit_scaledependent}
P_0(p, \zeta |p', \zeta') = \frac{1}{\sqrt{2 \pi \Delta V}} \left\{ \exp \left[ -\frac{(p - p')^2}{2 \Delta V} \right] - a' \exp \left[ -\frac{(p + p')^2}{2 \Delta V} \right] - b' \exp \left[ -\frac{(p - 2 H_2 + p')^2}{2 \Delta V} \right] \right\}, 
\end{equation}
Substituting Eq. \eqref{eq:P0p_transit_scaledependent} into two boundary conditions ($P_0(p = 0, \zeta | p', \zeta') = 0$ and Eq. \eqref{eq:transition_nonabsorbing_BC}), we obtain two linear equations
\begin{align}
&a' \cdot \exp \left[ -\frac{(p')^2}{2 \Delta V} \right] + b' \cdot  \exp \left[ -\frac{(p' - 2 H_2)^2}{2 \Delta V} \right] = \exp \left[ -\frac{(p')^2}{2 \Delta V} \right], \\
&a' \cdot \left\{ \frac{1}{\sqrt{2 \pi \Delta V}} \exp \left[ -\frac{(H_2 + p')^2}{2 \Delta V} \right] - \frac{\Delta H}{\Delta V} \text{erfc} \left( \frac{H_2 + p'}{\sqrt{2 \Delta V}} \right) \right\} + \notag \\
&b' \cdot \left\{ \frac{1}{\sqrt{2 \pi \Delta V}} \exp \left[ -\frac{(H_2 - p')^2}{2 \Delta V} \right] + \frac{\Delta H}{\Delta V} \text{erfc} \left( \frac{H_2 - p'}{\sqrt{2 \Delta V}} \right)\right\} = \notag \\
&\frac{1}{\sqrt{2 \pi \Delta V}} \exp \left[ -\frac{(H_2 - p')^2}{2 \Delta V} \right] - \frac{\Delta H}{\Delta V} \text{erfc} \left( \frac{H_2 - p'}{\sqrt{2 \Delta V}} \right), 
\end{align}
where $\text{erfc}()$ is the complementary error function. Given $p'$, $\Delta V$, and $\Delta H$, $a'$ and $b'$ can be solved analytically. The relative plastic contact area and the relative non-contact area are determined based on load equilibrium and probability conservation:
\begin{align}
A_{\text{pl}}^*(\zeta) &= \left[ \bar{p} - \int_0^{H_2} p P_0(p, \zeta) dp \right]/H_2, \label{eq:Apl_form_scale} \\
A_{\text{nc}}^*(\zeta) &= 1 - \int_0^{H_2} P_0(p, \zeta) dp - A_{\text{pl}}^*(\zeta). \label{eq:Anc_form_scale}
\end{align}
The contact pressure PDF at the target magnification $\zeta$ is solved sequentially by tracing the increasing sequence of $\zeta = [\zeta_0, \cdots, \zeta_i, \cdots, \zeta_n, \zeta]$ using the following compounded Chapman-Kolmogorov equation: 
\begin{align}
P(p, \zeta) = &\int_0^{H(\zeta_n)} P(p, \zeta|p_n, \zeta_n) \int_0^{H(\zeta_{n-1})} P(p_n, \zeta_n|p_{n-1}, \zeta_{n-1}) \cdots  \notag \\
&\int_0^{H(\zeta_0)} P(p_1, \zeta_1|p_0, \zeta_0) P(p_0, \zeta_0) dp_0 \cdots dp_{n-1} dp_n. 
\end{align}
The compounded Chapman-Kolmogorov equation is equivalent to solving Eqs. \eqref{eq:CK_vary_H_P0}, \eqref{eq:Apl_form_scale}, and \eqref{eq:Anc_form_scale} sequentially. To accurately capture the influence of $H(\zeta)$ on the elastoplastic contact solution, the magnification step, $\zeta_i - \zeta_{i-1}$, utilized in the compounded Chapman-Kolmogorov equation, must be sufficiently small. 

\section{Numerical implementation}
To solve $P_0(p, \zeta)$, $P_0(p, \zeta' \in [1, \zeta))$ must be known in advance. The magnification axis within the range of $[1, \zeta]$ is discretized using a logarithmic scale: $\log_{10}(\zeta_j) =  j \cdot \Delta \zeta$, $j = 0, \cdots, n_{\zeta}$, where $\Delta \zeta = \log_{10} (\zeta)/(n_{\zeta} - 1)$. The corresponding hardness and pressure variance are denoted by $H_j = H(\zeta_j)$ and $V_j = V(\zeta_j)$, respectively. Due to the limited memory of the computational facility, we restrict the span of $P_0(p, \zeta_j)$ on the pressure axis over a finite pressure range as follows:
\begin{itemize}
\item Linear elastic contact, $p \in [0, \bar{p} + 6 \sqrt{V_j}]$;

\item Elastoplastic contact with constant hardness, $p \in [0, \min(\bar{p} + 6 \sqrt{V_j}, H_0)]$;

\item Elastoplastic contact with scale-dependent hardness, $p \in [0, \min(\bar{p} + 6 \sqrt{V_j}, H_j)]$.

\end{itemize}
This finite pressure range is uniformly discretized with a mesh interval of $\Delta p$. At the mesh grid $(p_j, \zeta_j)$, the discretized solutions are denoted as follows: $P^{(j)}_i = P_0(p_i, \zeta_j)$, $P_{ik}^{(j, j-1)} = P_0(p_i, \zeta_j|p_k, \zeta_{j-1})$, $A_j^{*\text{pl}} = A_{\text{pl}}^*(\zeta_j)$, $A_j^{*\text{nc}} = A_{\text{nc}}^*(\zeta_j)$. 

The discretized form of Eq. \eqref{eq:CK_vary_H_P0} with $\zeta_j$, $j \geq 2$, can be expressed as follows: 
\begin{equation}\label{eq:CK_vary_H_P0_discrete}
P_i^{(j)} = \sum_{k = 1}^{n_p} K_{ik}^{(j, j-1)} P_k^{(j-1)} + P_0(p_i, \zeta_j|H_1, \zeta_{j-1}) A_{j-1}^{*\text{pl}}, 
\end{equation}
where $K_{ik}^{(j, j-1)} = P_0(p_i, \zeta_j|p_k, \zeta_{j-1}) \Delta p$ and $n_p$ denotes the number of pressure-grid points at a given magnification. According to Eq. \eqref{eq:P0p_transit_scaledependent}, the span of $P_0(p_i, \zeta_j|p_k, \zeta_{j-1})$ on the pressure grid is proportional to $\sqrt{\Delta V} = \sqrt{V_j - V_{j-1}}$. To accurately represent the shape of the transition probability, $\Delta p$ must be coupled with the magnification step $\Delta \zeta$ as follows:
\begin{itemize}
\item Linear elastic contact, $\Delta_p = \sqrt{\Delta V}/10$; 
\item Elastoplastic contact, $\Delta_p = \min(\sqrt{\Delta V}/10, \Delta p_0)$.
\end{itemize}
At low magnification, $\Delta V$ is extremely small, making $\Delta p$ closely linked to $\Delta \zeta$. As magnification increases, $\Delta V$ inevitably increases in accordance with the logarithmic scale discretization strategy. We anticipate that the pressure mesh interval increases monotonically as magnification increases until it converges at the highest magnification. The constant $\Delta p_0$ is used to prevent an unacceptably low mesh density on the pressure axis at high magnification. For elastoplastic contact, $\Delta p_0 = H_0/(n_{p0}-1)$, where $n_{p0}$ denotes the prescribed reference number of pressure-grid points used to define $\Delta p_0$. Since the pressure mesh interval is coupled with the magnification step, the pressure PDFs at neighboring magnifications are defined on different pressure grids. An additional mapping of the pressure PDF at $\zeta_{j-1}$ onto the pressure grid at $\zeta_j$ is implemented through linear interpolation prior to utilizing Eq. \eqref{eq:CK_vary_H_P0_discrete}. The discretized forms of Eqs. \eqref{eq:CK_vary_H_Apl} and \eqref{eq:CK_vary_H_Anc} are as follows:
\begin{align}
A_j^{*\text{pl}} = \left[ \bar{p} - \Delta p \sum_{i = 1}^{n_p} p_i P_i^{(j)} \right]/H_j, \\
A_j^{*\text{nc}} = 1 - \Delta p \sum_{i = 1}^{n_p} P_i^{(j)} - A_j^{*\text{pl}}. 
\end{align}
When $j = 1$, $P_i^{(j-1)}$ is a Dirac delta function. Depending on $\bar{p}$, we may have different $P_i^{(j)}$:
\begin{equation}
   P_i^{(j)} = \frac{1}{\sqrt{2 \pi \Delta V}} \left\{ \exp \left[ -\frac{(p_i - \bar{p})^2}{2 \Delta V} \right] - a' \exp \left[ -\frac{(p_i + \bar{p})^2}{2 \Delta V} \right] - b' \exp \left[ -\frac{(p_i - 2 H_j + \bar{p})^2}{2 \Delta V} \right] \right\} ~~~ \bar{p} < H_0, 
\end{equation}
\begin{equation}
   P_i^{(j)} = \frac{1}{\sqrt{2 \pi \Delta V}} \left\{ \exp \left[ -\frac{(p_i - H_0)^2}{2 \Delta V} \right] - a' \exp \left[ -\frac{(p_i + H_0)^2}{2 \Delta V} \right] - b' \exp \left[ -\frac{(p_i - 2 H_j + H_0)^2}{2 \Delta V} \right] \right\} ~~~ \bar{p} \geq H_0.
\end{equation}
In the work of \citet{lambert2025competition}, the numerical results are obtained from the diffusion equation. It involves solving a system of linear equations with a tri-diagonal matrix, which requires more computational time than the present model.  

\section{Results}

\begin{figure}[h!]
  \centering
  \includegraphics[width=16cm]{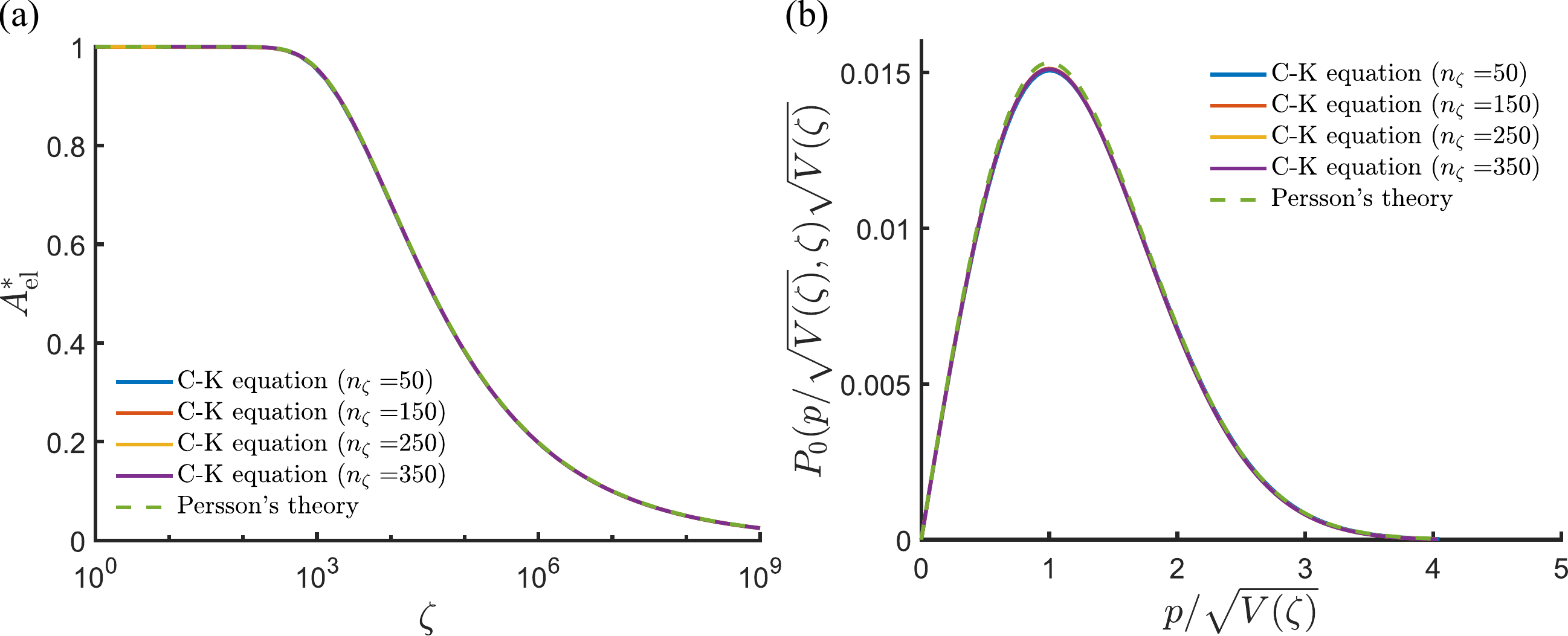}
  \caption{Convergence test of linear elastic contact solutions solved by Chapman-Kolmogorov (C-K) equation (solid lines) and Persson’s theory (dashed lines): (a) relative contact area $A^*$, (b) $P_0(p/\sqrt{V(\zeta)}, \zeta)$ at $\zeta = 1 \times 10^9$. Magnification axis is discretized by various $n_{\zeta}$, respectively. The mean pressure $\bar{p} = 1.5$ GPa, $H_0 = 3$ GPa, $C_0 = 1 \times 10^{-8}$ mm$^4$, $q_{\text{l}} = 2 \pi$ mm$^{-1}$, $H = 0.7$, $E = 200$ GPa, $\nu = 0.33$.}\label{fig:Fig_R1}
\end{figure}

\begin{table}[h!]
\caption{Convergence test of elastic and elastoplastic contact solutions. Elastic contact: mean absolute difference between numerical solution and Persson's theory over the investigated magnification range as shown in Fig. \ref{fig:Fig_R1}; Elastoplastic contact: mean absolute difference between numerical solutions with $n_{\zeta} = 50, 150, 250$, and numerical solutions with $n_{\zeta} = 350$ over the investigated magnification range as shown in Figs. \ref{fig:Fig_R2} and \ref{fig:Fig_R3}. }\label{tab:Tab_R1}
\vspace{1em}
\begin{tabular}{c| cc| ccc| ccc}
\hline
\multirow{2}{*}{$n_{\zeta}$} & \multicolumn{2}{c|}{Elastic contact}    & \multicolumn{3}{c|}{Elastoplastic contact $H_0$}                       & \multicolumn{3}{c}{Elastoplastic contact $H(\zeta)$}                  \\ 
                    & \multicolumn{1}{c}{$A^*$}        &  \multicolumn{1}{c|}{$P_0(p, \zeta)$}        & \multicolumn{1}{c}{$A_{\text{el}}^*$}         & \multicolumn{1}{c}{$A_{\text{pl}}^*$}         &  \multicolumn{1}{c|}{$P_0(p, \zeta)$}         & \multicolumn{1}{c}{$A_{\text{el}}^*$}         & \multicolumn{1}{c}{$A_{\text{pl}}^*$}         &  \multicolumn{1}{c}{$P_0(p, \zeta)$}        \\ \hline
50                  & \multicolumn{1}{c}{5.52e-4} & 3.09e-4  & \multicolumn{1}{c}{1.22e-04} & \multicolumn{1}{c}{2.84e-05} & 4.37e-04 & \multicolumn{1}{c}{1.49e-2} & \multicolumn{1}{c}{8.26e-04} & 2.74e-04 \\ \hline
150                 & \multicolumn{1}{c}{2.70e-4} & 2.04e-4  & \multicolumn{1}{c}{1.08e-04} & \multicolumn{1}{c}{2.42e-05} & 3.25e-04 & \multicolumn{1}{c}{7.42e-03} & \multicolumn{1}{c}{3.91e-04} & 1.83e-04 \\ \hline
250                 & \multicolumn{1}{c}{1.41e-4} & 1.02e-4  & \multicolumn{1}{c}{1.09e-04} & \multicolumn{1}{c}{2.63e-05} & 1.34e-04 & \multicolumn{1}{c}{2.80e-03} & \multicolumn{1}{c}{2.2e-04} & 8.98e-05 \\ \hline
350                 & \multicolumn{1}{c}{0}       & 4.62e-10 & \multicolumn{3}{c|}{N/A}                                                 & \multicolumn{3}{c}{N/A}                                                 \\ \hline
\end{tabular}
\end{table}
The convergence of the present model is explored first. When $H_0$ is significantly large, elastoplastic contact simplifies to linear elastic contact, for which the closed-form solutions are summarized in Appendix B. As shown in Fig. \ref{fig:Fig_R1}(a), the numerical results of $A^*(\zeta)$, predicted by the present model and Persson's theory, are nearly identical, regardless of the selected $n_{\zeta}$. Similarly, Fig. \ref{fig:Fig_R1}(b) shows that the pressure PDF $P_0(p, \zeta)$ at $\zeta = 1 \times 10^9$ predicted by the present model exhibits good agreement with the closed-form solution across the entire investigated pressure range, except for slight deviations at the peak value. As $n_{\zeta}$ increases from $50$ to $350$, the mean absolute differences of $A^*(\zeta)$ and $P_0(p, \zeta)$ over the investigated magnification ranges show a clear decreasing trend till vanishing at $n_{\zeta} = 350$, as indicated in Table \ref{tab:Tab_R1}. The difference in the pressure PDF can be attributed to two factors: (1) The Chapman-Kolmogorov equation and the diffusion equation in Persson's theory are not exactly equivalent \citep{xu2024persson}; (2) the numerical error that accumulates when solving the compounded Chapman-Kolmogorov equation.  

\begin{figure}[h!]
  \centering
  \includegraphics[width=16cm]{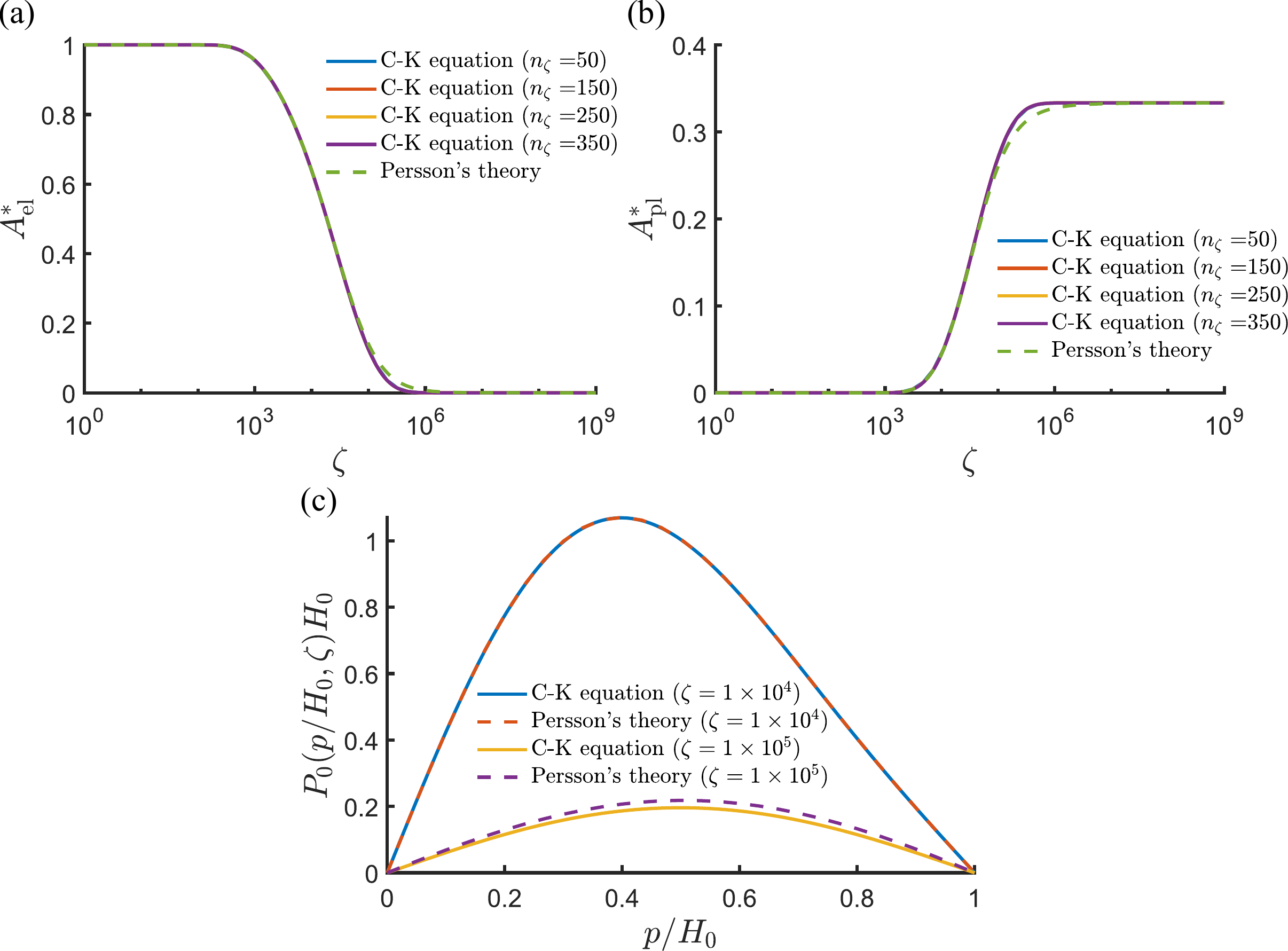}
  \caption{Convergence test of elastoplastic contact solutions with constant hardness solved by Chapman-Kolmogorov (C-K) equation (solid lines) and Persson’s theory (dashed lines): (a) relative elastic contact area $A_{\text{el}}^*$, (b) relative plastic contact area $A_{\text{pl}}^*$, (c) $P_0(p/H_0, \zeta)$. The magnification axis is discretized by various $n_{\zeta}$. The mean pressure $\bar{p} = 1.5$ GPa, $H_0 = 3$ GPa, $C_0 = 1 \times 10^{-8}$ mm$^4$, $q_{\text{l}} = 2 \pi$ mm$^{-1}$, $H = 0.7$, $E = 200$ GPa, $\nu = 0.33$, $\Delta p_0 = H_0/99$.}\label{fig:Fig_R2}
\end{figure}

For elastoplastic contact with finite, constant hardness $H_0$, \citet{Xu22} derived the closed-form solutions, which are summarized in Appendix B. As shown in Fig. \ref{fig:Fig_R2}(a-c) and Table \ref{tab:Tab_R1}, the numerical results of the present model demonstrate excellent convergence, yielding nearly identical results regardless of $n_{\zeta}$. At relatively low magnification ranges ($\zeta < 1 \times 10^5$), $A_{\text{el}}^*$, $A_{\text{pl}}^*$, and $P_0(p, \zeta)$ predicted by the present model align identically with those from Persson's theory. At higher magnification ranges ($\zeta \in [1 \times 10^5, 1 \times 10^6]$), where $A_{\text{el}}^*$ does not vanish, the present model slightly underestimates $P_0(p, \zeta)$ and $A_{\text{el}}^*$. This underestimation strongly suggests that the Chapman-Kolmogorov equation (Eq. \eqref{eq:CK_equ1}) and the diffusion equation (Eq. \eqref{eq:Diffusion_equ_P0}) are not strictly equivalent. That is why, when calculating the absolute mean difference in Table \ref{tab:Tab_R1}, we use the solution with the finest mesh as the reference solution rather than the solution of Persson's theory. 

\begin{figure}[h!]
  \centering
  \includegraphics[width=16cm]{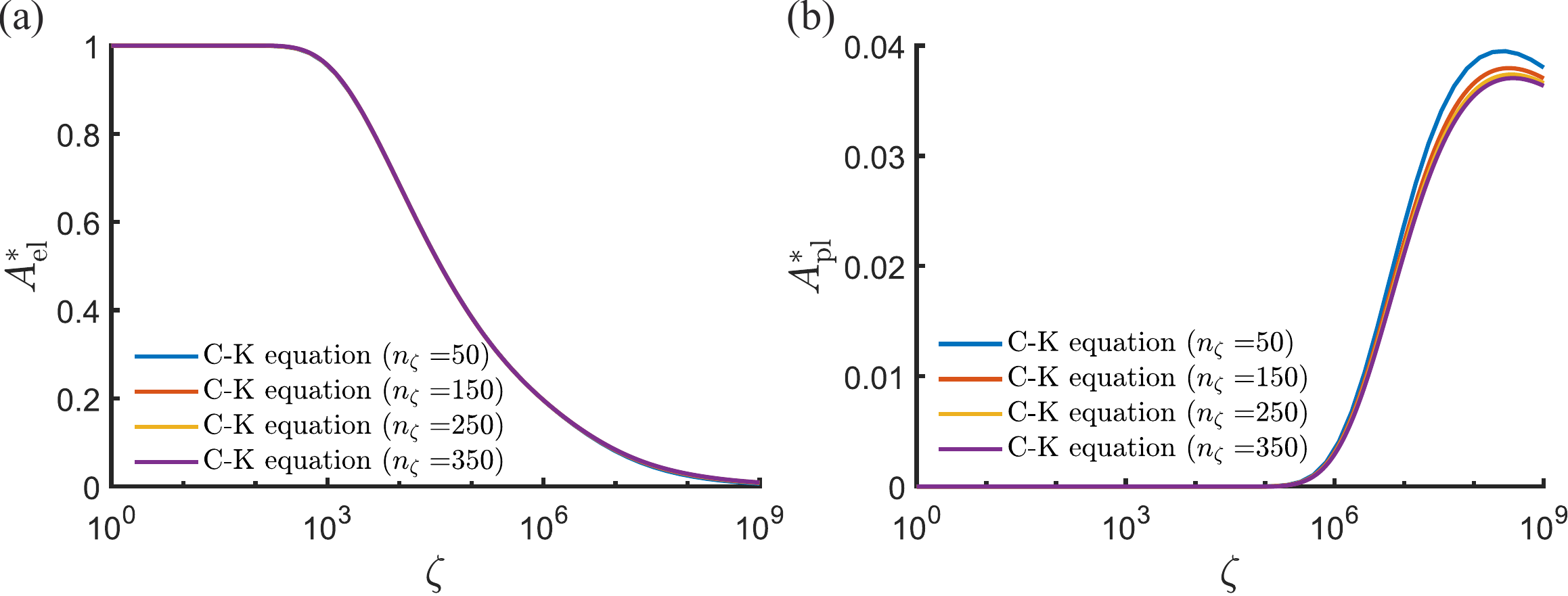}
  \caption{Convergence test of elastoplastic contact solutions with scale-dependent hardness solved by Chapman-Kolmogorov (C-K) equation (solid lines) and Persson’s theory (dashed lines): (a) relative elastic contact area $A_{\text{el}}^*$, (b) relative plastic contact area $A_{\text{pl}}^*$. The magnification axis is discretized by various $n_{\zeta}$. The mean pressure $\bar{p} = 1.5$ GPa, $H_0 = 3$ GPa, $n = 0.1$, $C_0 = 1 \times 10^{-8}$ mm$^4$, $q_{\text{l}} = 2 \pi$ mm$^{-1}$, $H = 0.7$, $E = 200$ GPa, $\nu = 0.33$, $\Delta p_0 = H_0/99$.}\label{fig:Fig_R3}
\end{figure}

In the present model, the scale-dependent hardness model proposed by \citet{lambert2025competition} below is adopted:
\begin{equation}
H(\zeta) = H_0 \zeta^n. 
\end{equation}
The convergence of the present model remains excellent for elastoplastic contact with scale-dependent hardness, $H(\zeta)$, as demonstrated in Fig. \ref{fig:Fig_R3} and Table \ref{tab:Tab_R1}. Figure \ref{fig:Fig_R3} illustrates an almost invariant $A_{\text{el}}^*$ against the selected $n_{\zeta}$, while a clear decreasing trend in the convergence of $A_{\text{pl}}^*$ is observed in Fig. \ref{fig:Fig_R3}(b). In summary, the present model is capable of generating a converged solution for all investigated $n_{\zeta}$. Without further notice, the following results solved by this work use the magnification step of $\Delta \zeta = 1/50$ on the logarithmic scale. 

\begin{figure}[h!]
  \centering
  \includegraphics[width=16cm]{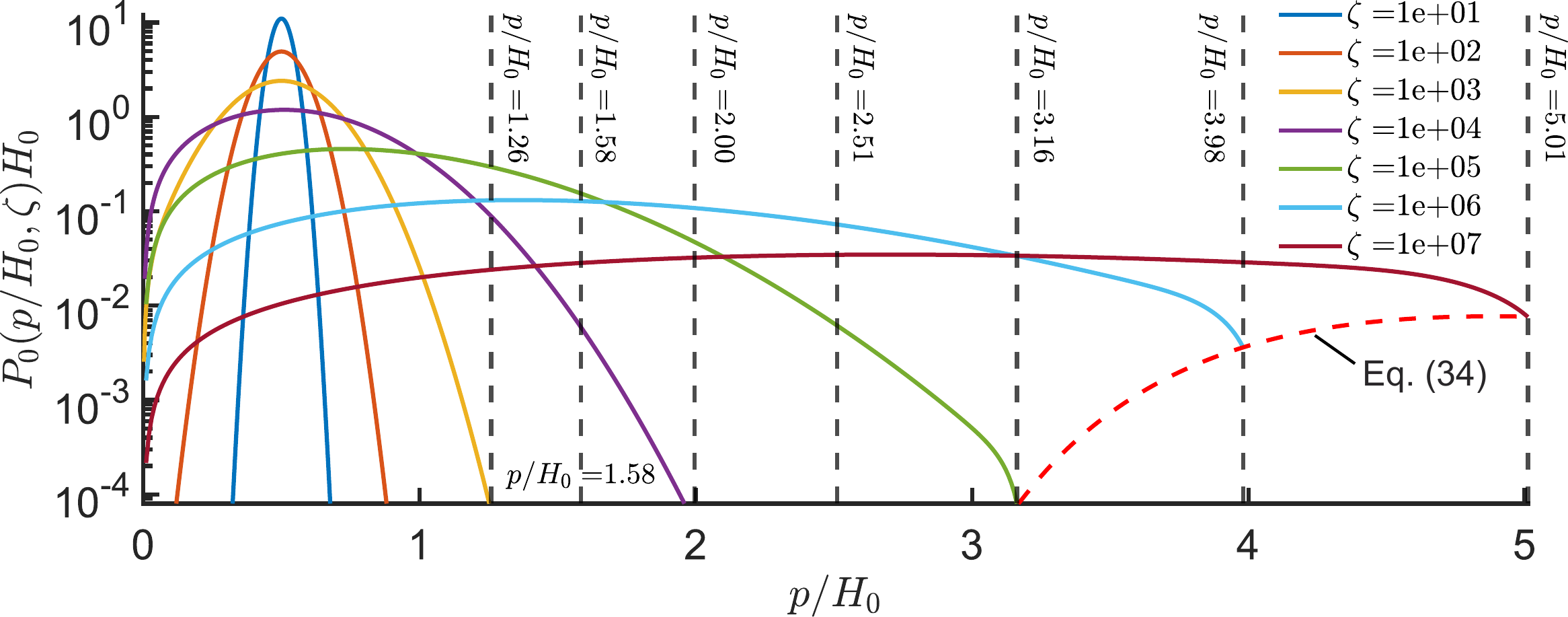}
  \caption{Evolution of $P_0(p/H_0, \zeta)$ with respect to $\zeta$. The mean pressure $\bar{p} = 1.5$ GPa, $H_0 = 3$ GPa, $n = 0.1$, $C_0 = 1 \times 10^{-8}$ mm$^4$, $q_{\text{l}} = 2 \pi$ mm$^{-1}$, $H = 0.7$, $E = 200$ GPa, $\nu = 0.33$, $\Delta p_0 = H_0/99$, $\Delta \zeta = 1/50$.}\label{fig:Fig_R4}
\end{figure}
Fig. \ref{fig:Fig_R4} demonstrates how the competition between $H(\zeta)$ (the upper limit of elastic contact pressure) and $V(\zeta)$ (redistribution of elastic contact pressure) influences $P_0(p, \zeta)$ at different magnifications. At low magnification ranges ($\zeta < 1 \times 10^5$), $P_0(p, \zeta)$ is broadened insignificantly over the pressure axis, resulting in a negligible plastic contact area. This can be confirmed by Fig. \ref{fig:Fig_R3}(b), where $A_{\text{pl}}^*$ begins to increase when $\zeta \geq 1 \times 10^5$. At higher magnification, $P_0(p, \zeta)$ accumulates at $p = H(\zeta)$. We can anticipate that the contact status gradually transitions from linear elastic to fully plastic status as $\zeta$ further increases. The evolution of $P_0(p, \zeta)$ with $\zeta$ qualitatively matches that solved by \citet{lambert2025competition} using the diffusion equation. 

\begin{figure}[h!]
  \centering
  \includegraphics[width=14cm]{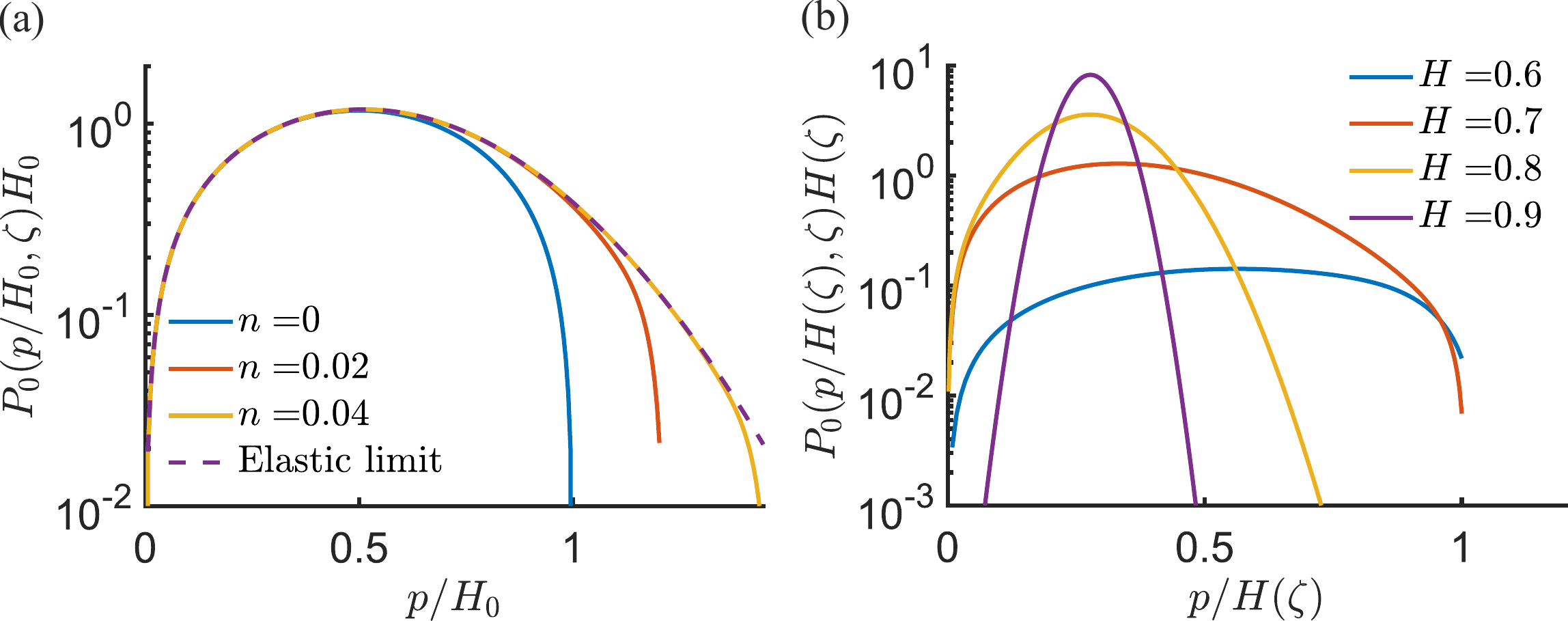}
  \caption{Evolution of $P_0(p/H_0, \zeta)$ with respect to (a) $n$ and (b) $H$. The mean pressure $\bar{p} = 1.5$ GPa, $H_0 = 3$ GPa, $C_0 = 1 \times 10^{-8}$ mm$^4$, $q_{\text{l}} = 2 \pi$ mm$^{-1}$, $E = 200$ GPa, $\nu = 0.33$, $\Delta p_0 = H_0/99$, $\Delta \zeta = 1/50$, $\zeta = 1 \times 10^4$, $H = 0.7$ in (a) and $n = 0.02$ in (b).}\label{fig:Fig_R5}
\end{figure}

The hardness exponent $n$ and the Hurst exponent $H$ significantly influence the elastoplastic contact solution. As $n$ increases, the increased hardness allows a larger fraction of the interface to undergo elastic deformation. Consequently, $P_0(p, \zeta)$ gradually transitions from an elastoplastic solution to a linear elastic solution, as illustrated in Fig. \ref{fig:Fig_R5}(a). It is expected that, when $n$ is sufficiently large, the error introduced by assuming fully elastic behavior throughout will be negligible. Thus, using a linear elastic model may be a good approximation for rough surface contact with a significant size effect, particularly when elastic contact is dominant. According to the boundary condition in Eq. \eqref{eq:transition_nonabsorbing_BC}, we can expect that $P_0(p= H(\zeta), \zeta)$ with non-zero $n$ decreases with increasing $n$. This anticipation is consistent with the numerical results shown in Fig. \ref{fig:Fig_R5}(a). Figure \ref{fig:Fig_R5}(b) shows that, as $H$ increases, $P_0(p, \zeta)$ transitions from a fully plastic solution to an elastic one. This conclusion is universal for rough surface contact, as $V(\zeta)$ decreases with increasing $H$ until it reaches zero when $H = 1$. Since $H(\zeta)$ is scale-dependent, there exists a competition between the redistribution of elastic contact pressure (governed by $V(\zeta)$) and the increase of the upper pressure limit (governed by $H(\zeta)$). As $H$ increases, the decrease of $V(\zeta)$ and its rate $dV/d\zeta$ progressively slow down the broadening of elastic contact pressure, while $H(\zeta)$ remains unchanged. Therefore, a smaller fraction of the interface undergoes plastic yielding at higher $H$. This competition has been thoroughly explored by \citet{lambert2025competition}, and a topographic yield parameter has been proposed. A new topographic parameter associated with the present model is proposed in the discussion section.

\begin{figure}[h!]
  \centering
  \includegraphics[width=16cm]{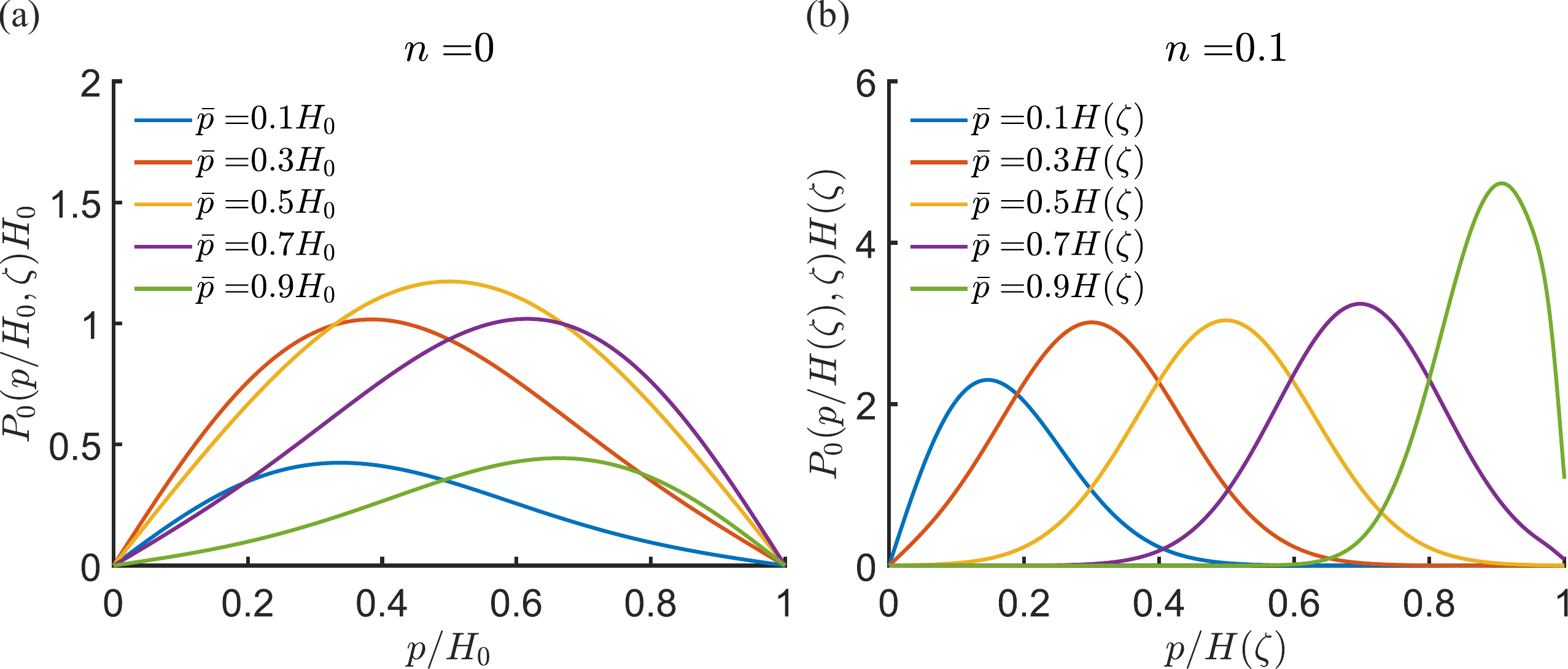}
  \caption{Evolution of $P_0(p/H_0, \zeta)$ with respect to $\bar{p}$ with (a) $n = 0$ and (b) $n = 0.1$. The hardness $H_0 = 3$ GPa, $C_0 = 1 \times 10^{-8}$ mm$^4$, $q_{\text{l}} = 2 \pi$ mm$^{-1}$, $E = 200$ GPa, $\nu = 0.33$, $\Delta p_0 = H_0/99$, $\Delta \zeta = 1/50$, $\zeta = 1 \times 10^4$.}\label{fig:Fig_R6}
\end{figure}

When $n = 0$, \citet{Xu22} and \citet{lambert2025competition} independently found that two elastic contact PDFs, under the normal loads $\bar{p}$ and $H_0 - \bar{p}$, respectively, are symmetric about $p = H_0/2$. This symmetry is confirmed in Fig. \ref{fig:Fig_R6}(a), while the scale-dependent hardness disrupts this symmetry (Fig. \ref{fig:Fig_R6}(b)). This asymmetry occurs because, when $n > 0$, the backflow of pressure probability from the plastic contact area to the elastic contact area is permitted (see Fig. \ref{fig:Fig_3}). Since backflow is more likely to occur between pressures with smaller differences, this leads to an accumulated probability peak in the close vicinity of $p = H(\zeta)$ (see the green curve in Fig. \ref{fig:Fig_R6}(b)). 

\begin{figure}[h!]
  \centering
  \includegraphics[width=16cm]{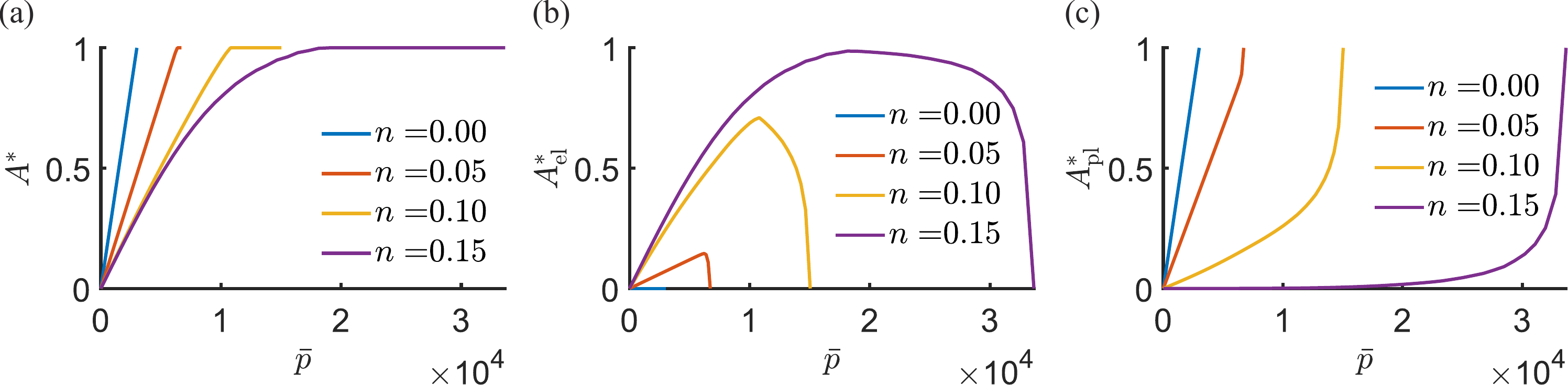}
  \caption{Variations of (a) relative contact area $A^* = A_{\text{el}}^* + A_{\text{pl}}^*$, (b) $A_{\text{el}}^*$, and (c) $A_{\text{pl}}^*$ with $\bar{p}$ and various $n$. The hardness $H_0 = 3$ GPa, $C_0 = 1 \times 10^{-8}$ mm$^4$, $q_{\text{l}} = 2 \pi$ mm$^{-1}$, $H = 0.7$, $E = 200$ GPa, $\nu = 0.33$, $\Delta p_0 = H_0/99$, $\Delta \zeta = 1/50$, $\zeta = 1 \times 10^7$.}\label{fig:Fig_R7}
\end{figure}

The influence of scale-dependent hardness on the area-to-load relationship is illustrated in Fig. \ref{fig:Fig_R7}(a), where two limits are clearly shown: fully plastic contact ($n = 0$) and linear elastic contact ($n = 0.15$). As $n$ increases, the interaction of rough surfaces enters the elastoplastic contact regime, where the elastic contact area begins to dominate in the low and medium load ranges, and the plastic contact area increases when nearly complete contact is achieved. When $n = 0.15$ is reached, the contact remains elastic until complete contact is attained. To maintain load equilibrium as $\bar{p}$ further increases beyond the initial complete contact, the elastic contact area transitions to the plastic contact area, while the relative contact area remains constant at $1$ (see the purple line in Fig. \ref{fig:Fig_R7}(b, c)). The common feature of all $A^*(\bar{p})$ curves is their linear portion at low load ranges, the slope of which can be characterized using $\kappa$.
\begin{equation}\label{eq:A_lowload}
A^* = \kappa \frac{\bar{p}}{E^* \sqrt{\langle |\nabla h |^2\rangle}},
\end{equation}
where $\sqrt{\langle |\nabla h |^2\rangle}$ is the root mean square gradient. It is clear that $\kappa$ is $H(\zeta)$-dependent. The slope $\kappa$ drops from $E^* \sqrt{\langle |\nabla h |^2\rangle}/H_0$ (fully plastic limit) to $\sqrt{8/\pi}$ (linear elastic limit) \citep{xu2024persson} as $n$ increases. 

\begin{figure}[h!]
  \centering
  \includegraphics[width=16cm]{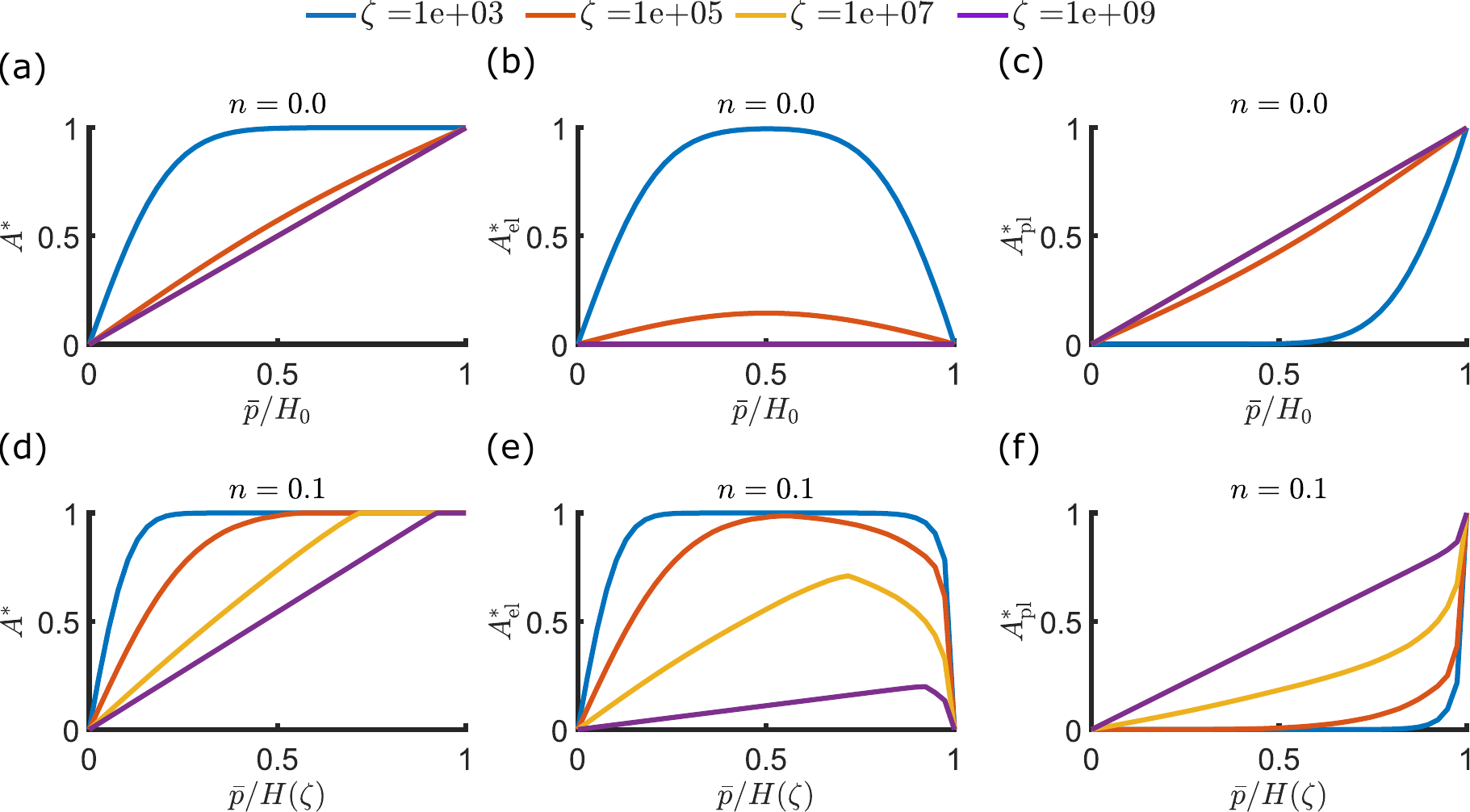}
  \caption{Variations of the relative contact area $A^* = A_{\text{el}}^* + A_{\text{pl}}^*$, $A_{\text{el}}^*$, and $A_{\text{pl}}^*$ with $\bar{p}$ with (a-c) $n = 0$ and (d-f) $n = 0.1$. The hardness $H_0 = 3$ GPa, $C_0 = 1 \times 10^{-8}$ mm$^4$, $q_{\text{l}} = 2 \pi$ mm$^{-1}$, $H = 0.7$, $E = 200$ GPa, $\nu = 0.33$, $\Delta p_0 = H_0/99$, $\Delta \zeta = 1/50$.}\label{fig:Fig_R8}
\end{figure}

Regardless of $n$, $P_0(p, \zeta)$ monotonically decreases with increasing $\zeta$. To maintain load equilibrium, plastic contact becomes dominant at higher $\zeta$. Therefore, we can anticipate that increasing $\zeta$ will result in a transition of the rough surface interaction from linear elastic contact to fully plastic contact. This is confirmed by Fig. \ref{fig:Fig_R8}. Similarly, all other mechanical and multiscale parameters (e.g., $E^*$, $H$, $q_{\text{l}}$, $C_0$) will influence the contact status. To quantify their joint effect on the contact status, a topographic yield parameter is proposed in the discussion section.   

\section{Discussion}

\subsection{Representative elastoplastic multiscale models}\label{subsec:model_discussion}

Two existing multiscale models, developed respectively by \citet{gao2006elastic} and \citet{jackson2006multi}, formulate the evolution of elastoplastic contact status with the scale similarly to that of the present model, even though these two models only consider size-independent plasticity. Those two models and the present model are all built upon the concept of ``protuberance-on-protuberance" originally proposed by \citet{archard1957elastic}. \citet{gao2006elastic} characterize the two-dimensional rough profile using the Weierstrass-Mandelbrot function, while \citet{jackson2006multi} approximates the three-dimensional rough surface using Fourier series. Thus, the protuberances in the Gao-Bower model and Jackson-Streator model are two-dimensional and three-dimensional periodic sinusoidal waviness with decreasing amplitudes at higher magnification, respectively. In the present model, the ``protuberance" is a random rough surface whose PSD occupies a limited bandwidth of $q \in [\zeta q_{\text{l}}, \zeta' q_{\text{l}}]$. Moreover, under complete elastic contact, the PDF of the contact pressure predicted by the present model is a Gaussian distribution, which implies that the rough surface height follows a Gaussian distribution \citep{xu2024persson}. Therefore, the evolution of the PDF of the rough surface height $P(h, \zeta)$ with $\zeta$ can be written in a step-wise fashion using the Chapman-Kolmogorov equation as follows:
\begin{equation}
P(h, \zeta) = \int_{-\infty}^{\infty} P(h, \zeta |h', \zeta') P(h', \zeta') dh'.
\end{equation}
The transition PDF is 
\begin{equation}
P(h, \zeta|h', \zeta') = \frac{1}{\sqrt{2 \pi \Delta V_h}} \exp \left( -\frac{(h - h')^2}{2 \Delta V_h} \right),
\end{equation} 
where $\Delta V_h$ is the change in the variance of the rough surface height when magnification transitions from $\zeta'$ to $\zeta$. Thus, the ``protuberances" in the present model are Gaussian rough surfaces with increasing variances at higher magnifications. 

Both the Gao-Bower model and the Jackson-Streator model utilized a hierarchical structure to solve the evolution of the relative contact area and contact pressure distribution in response to the stacking of sinusoidal components with an increasing wavenumber. Interestingly, \citet{gao2006elastic} used an integral equation to characterize the transfer of the PDF of the contact pressure between neighboring scales, just like the Chapman-Kolmogorov equation that we use in the present model. This equation was originally developed by \citet{ciavarella2000linear} to solve a linear elastic contact problem, where the transition probability is determined by solving the elastic sinusoidal waviness contact. In summary, the Gao-Bower and Jackson-Streator models are fundamentally the same as the present model, except for a different decomposition of the rough surface topography along the magnification axis. 

\citet{ciavarella2024new} modified Persson's elastoplastic contact theory to incorporate a scaling law of hardness developed by \citet{swadener2002correlation} based on the spherical indentation. The hardness is a constant within Persson's theory, which increases with the upper cut-off wavenumber. This is different from the present work, where the hardness progressively increases with the scale till the maximum magnification is reached. Additionally, a new plasticity index, the ratio of the mean contact pressure over the real area of contact to the hardness, was developed based on the small contact area approximation. The plasticity index, $\Psi \sim \zeta^{-H/2}$, implies that the deformation of the rough surface with a lower Hurst exponent behaves more plastic. This is consistent with the prediction of the present work. Since the hardness is relatively overestimated at each scale, we can expect that the model of Ciavarella predicts a more elastic contact than the present one. 

\subsection{Topographic yield parameter}
\citet{lambert2025competition} derived a topographic yield parameter that quantifies the competition between the redistribution of elastic contact pressure and the increase of hardness as $\zeta$ increases:
\begin{align}
w(\zeta) &= H^{''}(\zeta)/H'(\zeta) - V^{''}(\zeta)/V'(\zeta) \notag \\
         &= (n - 2 + 2H) \zeta^{-1}, \label{eq:w_LB}  
\end{align}
where prime and double prime superscripts represent the first and second derivatives with respect to magnification. The redistribution of the elastic contact pressure and the increase of hardness are expressed by $V^{''}(\zeta)/V'(\zeta)$ and $H^{''}(\zeta)/H'(\zeta)$. \citet{lambert2025competition} proposed that $n = -2 H + 2$ (according to Eq. \eqref{eq:w_LB}) and $n = -H + 2$ (found by \citet{brodsky2016constraints}) represent the elastic and fully plastic limits, respectively. 

Here, we propose an alternative form for $w(\zeta)$ based on our present model. The increase of hardness with $\zeta$ is quantified by its rate $H'(\zeta)$, which is more straightforward than $H^{''}(\zeta)/H'(\zeta)$. The redistribution of the elastic contact pressure is quantified using $d\sqrt{V}/d\zeta$, where $\sqrt{V}$ is proportional to the rate of change of the span of the pressure PDF. Considering a simple case where $q_r = q_l$ in $C(q)$, we have:
\begin{equation}
d\sqrt{V}/d\zeta = \frac{1}{2} E^* \sqrt{(1 - H) \pi C_0} ~q_{\text{l}}^{-H + 1} \zeta^{1 - 2H}/\sqrt{\zeta^{-2H + 2} - 1}. 
\end{equation}
If $\zeta \gg 1$, 
\begin{equation}
d\sqrt{V}/d\zeta = \frac{1}{2} E^* \sqrt{(1 - H) \pi C_0} ~q_{\text{l}}^{-H + 1} \zeta^{-H}. 
\end{equation}
Finally, a new topographic yield parameter is proposed as $w(\zeta) = H'(\zeta)/\left[ d \sqrt{V}/d\zeta \right]$: 
\begin{equation}\label{eq:w_Xu}
    w(\zeta) = \frac{2 H_0 n q_{\text{l}}^{H - 1}}{E^* \sqrt{(1 - H) \pi C_0}} \zeta^{n + H - 1}. 
\end{equation}
Unlike the $w(\zeta)$ proposed by \citet{lambert2025competition}, which occupies the entire real axis, $w(\zeta)$, defined in Eq. \eqref{eq:w_Xu}, is positive given $n > 0$ since $\zeta$ is finite. Comparing Eqs. \eqref{eq:w_LB} and \eqref{eq:w_Xu}, the new topographic yield parameter quantifies how other mechanical and topographic parameters, including $E^*$, $q_{\text{l}}$, and $C_0$, influence the transitions between elastic, elastoplastic, and fully plastic contact. 

\subsection{Elastoplastic contact status diagram}
As $w(\zeta)$ increases or decreases, the elastoplastic contact transitions to a more purely elastic state or to a fully plastic state, respectively. The scaling of Eq. \eqref{eq:w_Xu} suggests that the line $n + H - 1 = 0$ separates regimes where $w(\zeta)$ grows or decays with magnification. However, the numerical contours shown below indicate that this line should be interpreted only as a reference scaling line rather than as the practical elastic limit for the present model. Inspired by the misfit parameter originally proposed by \citet{lambert2025competition}, we introduce two parameters as follows:
\begin{align}
M_{\text{le}} =& \int_1^{\zeta} \left[ \bar{p} - H(\zeta') A_{\text{pl}}^*(\zeta') \right]^2 d\zeta'/ \left[ \bar{p}^2 \left( \zeta - 1 \right) \right] \times 100 \%, \\
M_{\text{fp}} =& \int_1^{\zeta} \left[ H(\zeta') A_{\text{pl}}^*(\zeta') \right]^2 d\zeta'/ \left[ \bar{p}^2 \left( \zeta - 1 \right) \right] \times 100 \%.  
\end{align}
The parameter $M_{\text{fp}} \in [0, 100] \%$ is equivalent to the mean ratio of the plastic load to the total load. As $M_{\text{fp}}$ increases towards $100 \%$, the elastoplastic solution converges to the fully plastic solution. Similarly, as $M_{\text{le}}$ increases from $0 \%$ to $100 \%$, the elastoplastic solution varies inversely from fully plastic to linear elastic. 

\begin{figure}[h!]
  \centering
  \includegraphics[width=16cm]{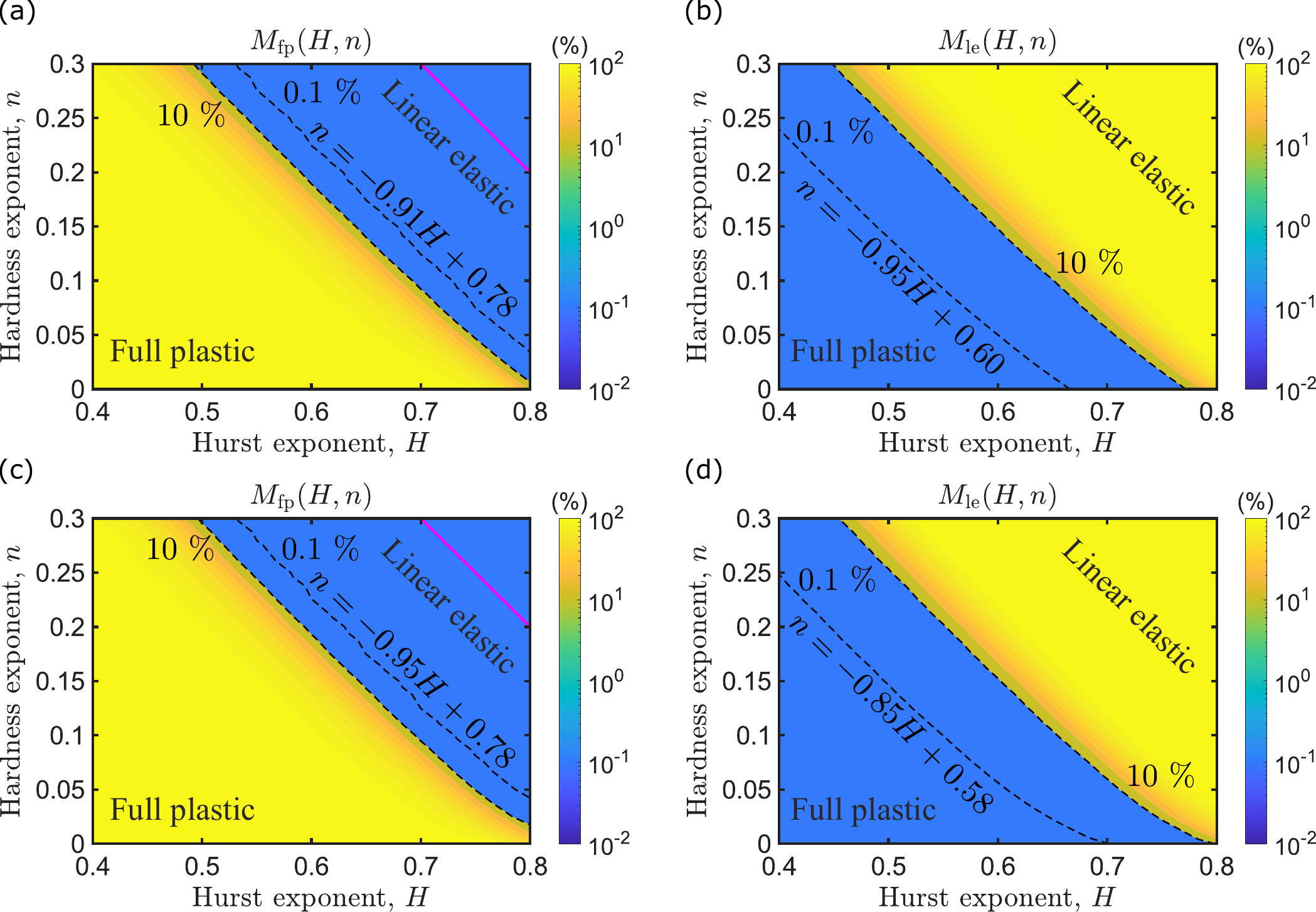}
  \caption{Contour plot of (a, c) $M_{\text{fp}}(H, n)$ and (b, d) $M_{\text{le}}(H, n)$: (a, b) $\bar{p} = 0.3 H_0$ and (c, d) $\bar{p} = 0.7 H_0$. The dashed lines are isolines of $0.1 \%$ and $10 \%$. The red line denotes the reference scaling line, $n = -H + 1$, inferred from Eq. \eqref{eq:w_Xu}; it is shown for comparison with the numerically calibrated elastic boundary. The hardness $H_0 = 3$ GPa, $C_0 = 1 \times 10^{-8}$ mm$^4$, $q_{\text{l}} = 2 \pi$ mm$^{-1}$, $E = 200$ GPa, $\nu = 0.33$, $\Delta p_0 = H_0/99$, $\Delta \zeta = 1/50$, $\zeta \in [1, 1 \times 10^7]$.}\label{fig:Fig_R9}
\end{figure}

Thanks to the high computational efficiency of the present model, the $(H,n)$ space used to generate the contours of $M_{\text{fp}}(H,n)$ and $M_{\text{le}}(H,n)$ is sampled more densely over the selected parameter ranges than those provided by \citet{lambert2025competition}; see their Fig. 10. The contour plot of $M_{\text{fp}}(H, n)$, as illustrated in Fig. \ref{fig:Fig_R9}(a), shows a distribution that is approximately uniform along $n = -H + c$. The entire contour is divided into three zones: linear elastic (bottom left triangle), elastoplastic (middle 45-degree narrow transition band), and fully plastic (upper right triangle). Although Eq. \eqref{eq:w_Xu} suggests the reference scaling line $n = -H + 1$, the contour results show that this line lies well outside the narrow elastoplastic transition band and therefore should not be interpreted as the practical elastic limit. It is important to note that as $M_{\text{fp}}(H, n)$ drops below $10 \%$, $M_{\text{fp}}(H, n)$ rapidly vanishes, and we define the elastic limit based on $M_{\text{fp}}(H, n) = 0.1 \%$. The linear fit of $n(H)$ that satisfies $M_{\text{fp}}(H, n) = 0.1 \%$ is $n = -0.91 H + 0.78$. The contour plot of $M_{\text{le}}(H, n)$, as shown in Fig. \ref{fig:Fig_R9}(b), shows the same rapid drop of $M_{\text{le}}(H, n)$ at $10 \%$. We define the plastic limit based on $M_{\text{le}}(H, n) = 0.1 \%$, where $n$ and $H$ satisfy $n = -0.95 H + 0.60$. The contours of $M_{\text{le}}(H, n)$ and $M_{\text{fp}}(H, n)$ shown in Fig. \ref{fig:Fig_R9} (a, b) are under the normal load of $\bar{p} = 0.3 H_0$. We found that increasing the normal load to $\bar{p} = 0.7 H_0$ does not significantly alter the elastic and plastic limits (Fig. \ref{fig:Fig_R9}(c, d)), although the isolines slightly flattened at the high $H$ ranges. 
 
\begin{figure}[h!]
  \centering
  \includegraphics[width=16cm]{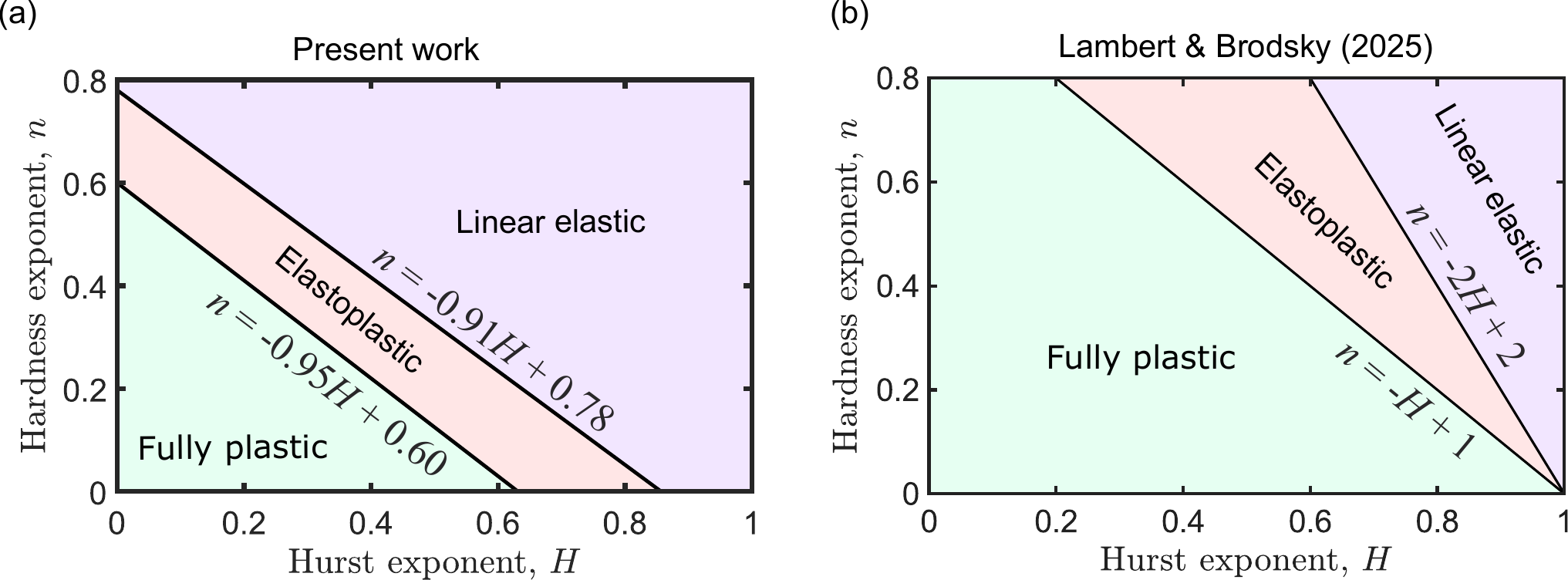}
  \caption{Graphical illustration of full plastic, elastoplastic and linear elastic zones: (a) present model, (b) work of \citet{lambert2025competition}.}\label{fig:Fig_R10}
\end{figure}

 Figure \ref{fig:Fig_R10}(a) illustrates three distinct zones divided by the elastic and plastic limits obtained from Fig. \ref{fig:Fig_R9}(a, b). For comparison, the zones divided by the limits found by \citet{lambert2025competition} are also provided in Fig. \ref{fig:Fig_R10}(b). The elastoplastic zones in both diagrams have different shapes and sizes. The difference is largely due to the fundamental distinction between the two models. The validity of both diagrams requires further validation through experiments or computational models. The present model (Fig. \ref{fig:Fig_R10}(a)) indicates that, for a common Hurst dimension of $H = 0.8$, the rough surface interface remains elastoplastically deformed with no size effect ($n = 0$) and easily enters the linear elastic limit with a relatively small increase in $n$. In contrast, the work of Lambert and Brodsky (Fig. \ref{fig:Fig_R10}(b)) suggests that the contact status remains fully plastic with no size effect ($n = 0$), and that a significant increase in $n$ is required to enter the elastoplastic and linear elastic ranges. However, it should be noted that the elastoplastic, fully elastic, and fully plastic ranges are distinguished based on the misfit parameters $M_{le}$ and $M_{lp}$, which quantify the mean relative difference, in percentage, between the prediction and its linear elastic and fully plastic limits, respectively, over a large span of magnifications $\zeta' \in [1, \zeta = 10^7]$. Therefore, the contact status identified in Fig. 13 should be applicable over a wide range of magnifications $\zeta \in [1, 10^7]$.

It should be noted that, even if the size effect results in a return to elastic behavior at small scales, the subsurface stress will approximately retain its plasticity, particularly at locations slightly beneath the interface where the strain gradient drops significantly. Therefore, the size effect may affect only the contact status at points very close to the boundary, where the local strain gradient is approximately of the same magnitude as that at the interface. Lastly, the immediate implication of the return to fully elastic behavior is ``asperity persistence,'' a classic phenomenon indicating that the local contact area can sustain a contact pressure much greater than the surface hardness. It is commonly attributed to work hardening \citep{childs1977persistence, inose2025asperity}. The present model may imply that the size effect can also be a possible mechanism for asperity persistence.

\subsection{Validation of the present work}
As far as the authors know, two physics-based rough surface contact models have successfully incorporated the size effect of plastic deformation \citep{song2016strain, venugopalan2019plastic}. However, those works do not discuss how hardness varies with scale, which makes the comparison nearly impossible. In the future study, the authors plan to develop the elastoplastic finite element contact model, which accounts for the size effect of the plastic deformation through strain gradient plasticity. The scale-dependent hardness can be estimated using an extra finite element model with a rigid flat-end punch indenting an elastoplastic half-space. The mean contact pressure once the contact area is fully yielded can be considered as the hardness. By continuously changing the size of the punch, the scale-dependent hardness can be obtained, which is essential for a fair comparison of the present model against the numerical models.

There is a lack of in-situ measurement of the metallic contact interface, particularly the real contact area and contact pressure distribution. Even though some metallic contact interfaces' geometrical properties have been measured using computer topography \citep{zhang2019experimental}, those tests are not designed to reveal the influence of the size effect on the elastoplastic contact. In the future study, the authors would like to overcome these obstacles by conducting the validation from a different perspective: According to the present model, as well as other similar elastoplastic contact models, the size effect of plastic deformation leads to an anti-intuitive phenomenon: the competition between the increasing hardness and increasing stress concentration at lower wavelength ranges leads to a fact that a more elastic behavior occurs at large wavelength ranges, leaving the plastic deformation accumulates within the low wavelength ranges. This phenomenon can be used to check the validity of the present model by measuring the deformed rough surface topography after a complete contact-separation cycle using the profilometer. The PSD of the rough surfaces before and after the contact should approximately agree at large wavelength ranges, but deviate obviously at the low wavelength range below a thresholding value. The deviation at a specific wavenumber should be correlated with $A_{\text{pl}}^*(\zeta)/A^*(\zeta)$.

\section{Conclusion}
In the present study, we developed a system of integral equations characterizing the evolution of the PDF of the elastic contact pressure, the relative plastic contact area, and the relative non-contact area with respect to magnification. This approach is applied to the context of elastoplastic contact involving scale-dependent hardness. A key advantage of our present model compared to the model of \citet{lambert2025competition}, which requires solving a system of linear equations with a tri-diagonal matrix, is the potential for a more computationally efficient solution process. As magnification increases, the elastic contact pressure becomes more uniformly distributed over a larger pressure range. The contact tends to be more elastic with a larger hardness exponent and a higher Hurst exponent. The symmetry of the PDF of the elastic contact pressure is broken due to the probability accumulation at $p = H(\zeta)$ induced by the probability backflow from the plastic contact area. The linear proportionality of the area-to-load relation drops from the fully plastic limit to the elastic limit as the hardness exponent increases. The present model, along with the Gao-Bower and Jackson-Streator models, can be conceptually traced back to the foundational framework established by the Archard model. While they share this common foundation, a key point of divergence lies in their methodologies for representing rough surface or profile topography. It is precisely these differing geometric representations that lead to the derivation of distinct governing equations for each model. We derive an asymptotic scaling criterion based on Eq. \eqref{eq:w_Xu} and then calibrate the practical elastic and plastic limits numerically. The calibrated boundaries, rather than the asymptotic line $n + H - 1 = 0$, provide the operative contact-status map for the present model. A contact status diagram with calibrated elastic and plastic limits, based on the numerical results of the present model, is provided for the quick identification of the contact status based on a finite number of mechanical and geometrical properties. The present work establishes, for the first time, a novel framework for applying the compounded Chapman-Kolmogorov equation to the analysis of rough surface contact. The integral equations developed herein, which characterize the evolution of interfacial properties with magnification, may offer insights into other multidisciplinary fields where multiscale roughness is a critical factor, such as the mechanics of geological faults and earthquake nucleation, as well as phenomena like contact electrification.

\section*{Acknowledgments}
YX thanks Robert Jackson, Michele Ciavarella, and Lucia Nicola for their inspiring discussions. This work was supported by the Fundamental Research Funds for the Opening Fund of the State Key Laboratory of Nonlinear Mechanics (LNM202507), the National Natural Science Foundation of China (No. 12172358), the Natural Science Foundation of Anhui Province (2508085ME101), and the Fundamental Research Funds for the Central Universities (JZ2025HGTG0298).

\begin{spacing}{1} 
    \bibliographystyle{cas-model2-names}
	\bibliography{ref}
\end{spacing}

\appendix 
\setcounter{figure}{0}
\setcounter{equation}{0}
\renewcommand{\thefigure}{A.\arabic{figure}}
\renewcommand{\theequation}{A.\arabic{equation}}
\renewcommand{\thetable}{A.\arabic{table}}

\section*{Appendix A. Boundary condition}
In this appendix we derive an approximate boundary condition for $P_0(p = H_2, \zeta|p', \zeta')$ within the small-step formulation introduced in Eq. \eqref{eq:Diffusion_equ_transition_1}, namely $dV/d\zeta \approx \Delta V/\Delta \zeta$ and $dH/d\zeta \approx \Delta H/\Delta \zeta$ over the interval $[\zeta', \zeta]$. Applying $\int_0^{H_2}  \cdots dp$ to both sides of Eq. \eqref{eq:Diffusion_equ_transition_1} and considering the identity that $\int_0^{H_2} P_0(p, \zeta|p', \zeta') dp = 1 - A_{\text{pl}}^*(\zeta) \bigg|_{p', \zeta'} - A_{\text{nc}}^*(\zeta) \bigg|_{p', \zeta'}$ results in  
\begin{equation}\label{eq:A1}
-\frac{\partial}{\partial \zeta} A_{\text{pl}}^*(\zeta)\bigg|_{p', \zeta'}-\frac{\partial}{\partial \zeta} A_{\text{nc}}^*(\zeta)\bigg|_{p', \zeta'} = \frac{1}{2} \frac{\Delta V}{\Delta \zeta} \frac{\partial}{\partial p} P_0(p, \zeta|p', \zeta')\bigg|_0^{H_2} + \frac{\Delta H}{\Delta \zeta} P_0(H_2, \zeta|p', \zeta').
\end{equation}
Notice that the second term on the right side of Eq. \eqref{eq:A1} arises from differentiation of the upper limit of integration. Because the ``leakage" of $P_0(p, \zeta|p', \zeta')$ with respect to $\zeta$ at $p = 0$ directly influences $A_{\text{nc}}^*(\zeta)\bigg|_{p', \zeta'}$, we can approximate the rate of change of $A_{\text{nc}}^*(\zeta)\bigg|_{p', \zeta'}$ with $\zeta$ by 
\[
\frac{\partial}{\partial \zeta} A_{\text{nc}}^*(\zeta)\bigg|_{p', \zeta'} = \frac{1}{2} \frac{\Delta V}{\Delta \zeta} \frac{\partial}{\partial p} P_0(p, \zeta|p', \zeta')\bigg|_0.
\]
Similarly, $A_{\text{pl}}^*(\zeta)\bigg|_{p', \zeta'}$ can be formulated as
\begin{equation}\label{eq:leakage_rate_plastic}
\frac{\partial}{\partial \zeta} A_{\text{pl}}^*(\zeta)\bigg|_{p', \zeta'} = -\frac{1}{2} \frac{\Delta V}{\Delta \zeta} \frac{\partial}{\partial p} P_0(p, \zeta|p', \zeta')\bigg|_{H_2} - \frac{\Delta H}{\Delta \zeta} P_0(H_2, \zeta|p', \zeta'). 
\end{equation}
Applying $\int_0^{H_2} p \cdots dp$ to both sides of Eq. \eqref{eq:Diffusion_equ_transition_1},  
\begin{align}\label{eq:load_equilibrium_transition}
&-H_2 \frac{\partial}{\partial \zeta} A_{\text{pl}}^*(\zeta)\bigg|_{p', \zeta'} - \frac{\Delta H}{\Delta \zeta} A_{\text{pl}}^*(\zeta)\bigg|_{p', \zeta'} - \frac{\Delta H}{\Delta \zeta} H_2 P_0(H_2, \zeta|p', \zeta') = \notag \\
&\frac{1}{2} \frac{\Delta V}{\Delta \zeta} \left[ H_2 \frac{\partial}{\partial p} P_0(p, \zeta|p', \zeta')\bigg|_{H_2} - P_0(p, \zeta|p', \zeta')\bigg|_0^{H_2} \right]. 
\end{align}
Substituting Eq. \eqref{eq:leakage_rate_plastic} and $P_0(p = 0, \zeta|p', \zeta') = 0$ into Eq. \eqref{eq:load_equilibrium_transition}, we finally obtain the boundary condition 
\begin{equation}\label{eq:BC_scale_dependent}
P_0(p=H_2, \zeta|p', \zeta') = \frac{2 \Delta H}{\Delta V} A_{\text{pl}}^*(\zeta)\bigg|_{p', \zeta'}.
\end{equation}
This boundary condition is fundamentally the same as Eq. (16) in \cite{lambert2025competition}. If hardness is constant, i.e., $\Delta H = 0$, Eq. \eqref{eq:BC_scale_dependent} degrades to the absorbing boundary condition. 

\setcounter{figure}{0}
\setcounter{equation}{0}
\renewcommand{\thefigure}{B.\arabic{figure}}
\renewcommand{\theequation}{B.\arabic{equation}}
\renewcommand{\thetable}{B.\arabic{table}}

\section*{Appendix B. Summary of solutions of Persson's theory of elastoplastic contact}

As $H_0$ is significantly large, elastoplastic contact transitions to linear elastic contact, for which the following closed-form solutions are derived using Persson's elastic contact theory:
\begin{align}
P_0(p, \zeta) &= \frac{1}{\sqrt{2 \pi V(\zeta)}} \left( \exp\left[ -\frac{(p - \bar{p})^2}{2 V(\zeta)} \right] - \exp\left[ -\frac{(p + \bar{p})^2}{2 V(\zeta)} \right] \right), \label{eq:pPDF_nominallyflat} \\ 
A^*(\zeta) &= \text{erf}\left( \bar{p}/\sqrt{2 V(\zeta)} \right). \label{eq:area_nominallyflat}
\end{align}

For elastoplastic contact with finite, constant hardness $H_0$, \citet{Xu22} derived the following closed-form solutions:
\begin{equation}\label{eq:Persson_elastoplastic}
P_0(p, \zeta) =  \frac{1}{\sqrt{2 \pi V(\zeta)}} \left\{ \exp\left[ -\frac{(p - \bar{p})^2}{2 V(\zeta)} \right] - a \cdot \exp\left[ -\frac{(p + \bar{p})^2}{2 V(\zeta)} \right] - b \cdot \exp\left[ -\frac{(p - 2 H_0 + \bar{p})^2}{2 V(\zeta)} \right] \right\},
\end{equation}
where the unknowns, $a$ and $b$, are
\begin{align}
a &= \left\{ 1 - \exp[-2(H_0 - \bar{p}) H_0/V(\zeta) ]\right\}/\left[ 1 - \exp(-2H_0^2/V(\zeta))\right], \label{eq:a_Persson}\\
b &= 1 - a \cdot \text{exp}\left(-2 \bar{p} H_0/V(\zeta) \right). 
\end{align}
The relative elastic and plastic contact areas, $A_{\text{el}}^*$ and $A_{\text{pl}}^*$, are 
\begin{align}
A_{\text{el}}^*(\zeta) &= \frac{1 + b}{2}\text{erf}\left( \frac{H_0 - \bar{p}}{\sqrt{2 V(\zeta)}} \right) + \frac{1 + a}{2} \text{erf} \left( \frac{\bar{p}}{\sqrt{2 V(\zeta)}} \right) - \frac{a}{2}\text{erf}\left( \frac{H_0 + \bar{p}}{\sqrt{2 V(\zeta)}} \right) + \frac{b}{2}\text{erf}\left( \frac{\bar{p} - 2H_0}{\sqrt{2 V(\zeta)}} \right), \label{E:final_Ael_elastoplastic} \\
A_{\text{pl}}^*(\zeta) &= -\frac{1 + b}{2}\text{erf}\left( \frac{H_0 - \bar{p}}{\sqrt{2 V(\zeta)}} \right) + \frac{a}{2} \text{erf} \left( \frac{H_0 + \bar{p}}{\sqrt{2 V(\zeta)}} \right) + \frac{1}{2}(1 - a + b).\label{E:final_Apl_elastoplastic}
\end{align}

\end{document}